\documentclass[12pt]{article}
\usepackage{setspace}
\usepackage[ruled,vlined,linesnumbered]{algorithm2e}

\usepackage[normalem]{ulem}
\usepackage[utf8]{inputenc}
\usepackage{amsmath, amsfonts}
\usepackage[margin=0.9in]{geometry}

\usepackage[noend]{algpseudocode}

\usepackage{graphics, graphicx}
\usepackage{authblk}
\usepackage{xcolor}
\usepackage{hyperref}

\usepackage{multirow}
\usepackage{caption}
\usepackage{booktabs}
\usepackage{natbib}
\usepackage{makecell} 
\usepackage{longtable}
\usepackage{subcaption} 
\usepackage{float}
\usepackage{pdflscape}
\usepackage{hyperref}
\usepackage{bm}
\usepackage{placeins}

\title{Bayesian Multi-Group Functional Factor Models with Parameter-Expanded Cumulative Shrinkage Priors}

\author{Xuanye Dai\thanks{Department of Statistics, Computer Science, Applications, University of Florence, Italy}, Anna Gottard$^{*}$, Michele Guindani\thanks{Department of Biostatistics, UCLA, Los Angeles, CA, USA}, and \\Marina Vannucci\thanks{Department of Statistics, Rice University, Houston, TX, USA}}

\date{ }
\begin{document}

\maketitle

\begin{abstract}
Functional data consist of trajectories observed over a continuous domain, such as time, space, or wavelength. Here we consider curves observed on different groups of subjects and propose a Bayesian multi-group functional factor analysis framework that jointly models the data via an explicit decomposition into group-specific mean functions and latent components that capture both common and distinct latent structures across the groups. We represent these functional components as linear combinations of a common set of B-spline bases, achieving a low-rank representation of the latent factors. We further impose a parameter-expanded cumulative shrinkage process prior on the factor loadings, which induces increasing shrinkage and automatically selects the number of active shared and group-specific factors.  
We evaluate the model's performance through simulation studies and show that the model accurately recovers the number of underlying factors and effectively distinguishes variations in functional observations driven by shared versus group-specific complex structures under various scenarios. For real data analysis, we apply the model to EEG data on alcoholic and healthy subjects and identify shared latent factors, that capture canonical characteristic components of the EEG curves, along with group-specific factors that reveal specific neural activity patterns. 
\end{abstract}

%---------------------------------------------------
\section{Introduction}

Functional data consist of trajectories observed over a continuous domain, such as time, space, or wavelength, and are naturally modeled as infinite-dimensional objects in a functional space \citep{RamsayandSilverman2005}. This type of data arises in a wide variety of scientific fields, including neuroscience,  public health, chemometrics, and econometrics, among others. In many modern studies, functional data are collected across multiple related studies, sites, experimental groups or conditions. 
For example, in the application of this paper, we consider data from a neuroscience study that involves electroencephalogram (EEG) data measured on alcoholics and healthy controls. 

Since functional data are inherently infinite-dimensional, dimension reduction is crucial for effective modeling and analysis. A widely used approach is functional principal component analysis (fPCA) which projects smooth trajectories on a finite set of orthonormal eigenfunctions, leading to functional principal component (fPC) scores as a parsimonious representation of the data. 
Early theoretical developments focused on fully observed functional data \citep{Dauxoisetal1982}, with extensions to densely and sparsely observed trajectories \citep{Castroetal1986, Shietal1996}. Beyond the classical setting, fPCA has been generalized to more complex data structures, including multilevel approaches for within- and between-subject variation in hierarchical functional data \citep{Di2009multilevel}, longitudinal analyses for multiple time points \citep{Grevenetal2011}, spatio-temporal fPCA for data with spatial and temporal dependencies \citep{Krzysko2024spatio}, and filtrated common fPCA for multi-group functional data, to separate shared and group-specific variation \citep{Jiao2024filtrated}. Bayesian approaches offer a natural way to extend these models, by accommodating hierarchical structures, prior specifications, and direct uncertainty quantification \citep{Crainiceanuetal2010}. \cite{SuarezandGhosal2017} proposed a Bayesian fPCA that models the covariance function via an approximate spectral decomposition. Each eigenfunction is represented as a linear combination of a finite set of basis functions, with the number of components and basis functions jointly selected for interpretability. \cite{Shamshoianetal2022} further extended this framework to handle longitudinal and multidimensional functional data, while \cite{margaritella2021parameter} introduced a Bayesian nonparametric model for clustering fPC scores to capture spatiotemporal brain patterns. 

Bayesian fPCA can be naturally formulated as a latent factor model, with Bayesian nonparametric priors on the factor components. This formulation allows computationally convenient estimation and direct inference on the model components. \cite{Montagnaetal2012} proposed a Bayesian latent factor regression model for functional data, in which each subject’s curve is represented as a linear combination of basis functions and a latent factor model is imposed on the coefficients of the basis expansion. The authors employed the multiplicative gamma process (MGP) prior of \cite{BhattacharyaandDunson2011} on the factor loadings. This shrinkage prior allows many of the loadings to be close to zero, thus inducing effective basis selection. \cite{KowalandCanale2023} proposed a semiparametric functional factor model incorporating an alternative prior construction for rank selection. Specifically, they extended the cumulative shrinkage process (CUSP) prior of \cite{Legramantietal2020}, a prior that enforces increasing shrinkage on successive columns of the factor loading matrix to eliminate redundant factors. 
\cite{Bolandetal2023} also considered a Bayesian fPCA framework as a latent factor model with B-spline basis expansions and imposed a modified MGP prior on the variance components of the factor loading matrix, building on the approaches of \cite{Montagnaetal2012}, \cite{SuarezandGhosal2017}, and \cite{Shamshoianetal2022}. Even though widely adopted, MGP priors are sensitive to hyperparameter selection and increasing shrinkage can be achieved only for specific settings of the hyperparameters, as shown in \cite{Durante2017}. 

While functional latent factor models are effective for single-group data, they are not designed to address multi-group settings, which often involve both common patterns shared across groups and group-specific variation. Multi-study latent factor models for independent (ie, non-functional) data, which jointly decompose variation into shared and study-specific latent factors,
have been proposed by \cite{DeVitoetal2019} and \cite{DeVitoetal2021}.
This framework has inspired several extensions aimed at enhancing flexibility and identifiability. For instance, 
\cite{Grabskietal2023} developed a framework that enables partial sharing of latent factors across arbitrary subsets of studies, and \cite{Chandraetal2024} proposed Subspace Factor Analysis, which mitigates information switching between shared and study-specific components. Also, \cite{BortolatoandCanale2026} investigated shrinkage priors for latent factors that flexibly capture complex structures, ranging from the absence of study-specific factors to factors shared within subgroups or across subsets of studies. Most recently, \cite{Maurietal2025+} improved upon the identifiability issues of these models by using spectral decompositions to estimate shared and study-specific residual structures. 
These models, however, cannot be directly applied to functional data recorded on dense, ordered grids.

In this paper, we extend the latent factor Bayesian regression model formulation of \cite{Bolandetal2023} to multi-group functional data. By adapting the latent factor framework of \cite{DeVitoetal2021} to functional data, we model the underlying functional trajectories through a low-rank latent factor decomposition with shared and group-specific structures, and centered around group-specific mean functions.  The shared structure is expected to capture variation common across all groups, while the group-specific structure is expected to reveal group-specific deviations. We represent these functional trajectories as linear combinations of a common set of cubic B-spline basis functions and promote model parsimony by placing a parameter-expanded cumulative shrinkage process prior on the factor loadings. This prior construction induces increasing shrinkage on higher-index components, effectively pruning redundant dimensions. 
Posterior inference is carried out via a Gibbs sampler, coupled with a post-processing strategy that addresses identifiability issues by first aligning posterior draws in the parameter space and then identifying interpretable loading structures in the covariance space. 
We evaluate the model’s performance through simulation studies under various latent structures and demonstrate that the model accurately recovers the number of underlying factors and effectively distinguishes shared from group-specific variation. We further demonstrate the performance of our proposed model with an application to EEG data on alcoholic and healthy subjects, where we identify shared latent factors that capture canonical characteristic components of the EEG curves which are common to both groups of subjects, along with group-specific factors that reveal group-specific neural activity patterns.

The remainder of this article is organized as follows. Section~\ref{sec:Methods} presents the model, the prior construction, and a discussion of posterior inference. Section~\ref{sec:Simulation} contains simulation studies and a sensitivity analysis, and Section~\ref{sec:Application} demonstrates the application of our proposed methodology to a real EEG dataset. Finally, Section~\ref{sec:Conclusion} concludes the paper.

%---------------------------------------------------
\section{Methods}
\label{sec:Methods}
The following sections detail the structure of the proposed Bayesian multi-group functional factor model (BMGFFM). Throughout the paper, we use capital letters to indicate matrices and bold small letters for vectors. 

\subsection{Bayesian Multi-Group Functional Factor Model}
\label{Model Specification}

Let $\boldsymbol{y}_{is} =(y_{is}(t_{1}), \ldots, y_{is}(t_{T}))^{\top}$ denote the observed curve for subject $i$ in group $s$, with $i=1,\ldots, n_{s}$ and $s=1,\ldots, S$. We model the data as noisy realizations of a smooth underlying function with an additive measurement error as
\begin{equation}
\label{eq:model}
y_{is}(t)=f_{is}(t)+\epsilon_{is}(t), \qquad t\in\{t_1,\ldots,t_T\},
\end{equation}
with $\epsilon_{is}(t) \sim \mathcal{N}(0, \sigma_{\epsilon_{s}}^{2})$.
We assume that the underlying functional trajectory, $f_{is}(t)$, can be modeled through a low-rank decomposition involving a group-specific mean function, $\mu_s(t)$, together with $L$ shared latent components, $\widetilde{\lambda}_{l}(t)$, $l=1,\ldots, L$, and $K_s$ group-specific latent components, $\widetilde{\phi}_{k}^{s}(t)$, $k=1,\ldots, K_s$, as \begin{equation}
{f}_{is}(t) = \mu_{s}(t) + \sum_{l=1}^{L}\eta_{ils}\widetilde{\lambda}_{l}(t) + \sum_{k=1}^{K_s}\rho_{iks}\widetilde{\phi}^{s}_{k}(t),
\label{f_is}
\end{equation}
with $\eta_{ils}$ the scores associated with the shared components and $\rho_{iks}$ the scores associated with the group-specific components.  We further represent the smooth components as linear combinations of the same set of $R$ B-spline basis functions, $(b_{1}(t),\ldots,b_{R}(t))$, as 
\begin{eqnarray*}
\mu_{s}(t)=\sum_{r=1}^{R}\beta_{rs}b_{r}(t),\\
\widetilde{\lambda}_{l}(t)=\sum_{r=1}^{R}\lambda_{rl}b_{r}(t),\\
\widetilde{\phi}_{k}^{s}(t)=\sum_{r=1}^{R}\phi_{rk}^{s}b_{r}(t),
\end{eqnarray*}
with $\beta_{rs}$, $\lambda_{rl}$ and $\phi_{rk}^{s}$ denoting B-spline coefficients, and with $L+K_{s} \ll R,$ for all $ s=1,\ldots,S$. Using the basis expansion, the full representation of the underlying smooth function 
is ${f}_{is}(t)=\sum_{r=1}^{R}\left(\beta_{rs} + \sum_{l=1}^{L}\eta_{il s}\lambda_{rl}  + \sum_{k=1}^{K_{s}} \rho_{iks}\phi_{rk}^{s} \right)b_{r}(t)$, and model \eqref{eq:model} can be expressed in matrix form as
\begin{equation}
    \boldsymbol{y}_{is} = \boldsymbol{f}_{is} + \boldsymbol{\epsilon}_{is} = B(\boldsymbol{\beta}_{s} + \Lambda\boldsymbol{\eta}_{is} + \Phi_{s}\boldsymbol{\rho}_{is}) + \boldsymbol\epsilon_{is},   
\label{multi-group model}
\end{equation}
with $\boldsymbol{f}_{is}=(f_{is}(t_{1}), \ldots, f_{is}(t_{T}))^{\top}$, $\boldsymbol\epsilon_{is}=(\epsilon_{is}(t_{1}),\ldots,\epsilon_{is}(t_{T}))^{\top}$, and
 $B=(\boldsymbol{b}_{1},\ldots,\boldsymbol{b}_{R})$ denoting the $T \times R$ matrix of B-spline basis functions with $\boldsymbol{b}_{r}=(b_{r}(t_{1}),\ldots,b_{r}(t_{T}))^{\top}$. 
In the factor model formulation \eqref{multi-group model}, the $R \times 1$ vector $\boldsymbol{\beta}_{s}=(\beta_{1s}, \ldots, \beta_{Rs})^{\top}$ denotes the group-specific mean coefficients, $\boldsymbol{\eta}_{is} =(\eta_{i1s},\ldots,\eta_{iLs})^{\top}$ is the $L \times 1$ vector of shared latent factors and the $R \times L$ matrix $\Lambda=(\boldsymbol{\lambda}_{1},\ldots,\boldsymbol{\lambda}_{L})$, with $\boldsymbol{\lambda}_{l}=(\lambda_{1l}, \ldots, \lambda_{Rl})^{\top}$, denotes the shared factor loadings. Moreover,  $\boldsymbol{\rho}_{is}=(\rho_{i1s},\ldots, \rho_{iK_{s}s})^{\top}$ 
is the $K_s \times 1$ vector of group-specific latent factors and the $R \times K_s$ matrix $\Phi_{s}=(\boldsymbol{\phi}^{s}_{1},\ldots,\boldsymbol{\phi}^{s}_{K_s})$, with $\boldsymbol{\phi}_{k}^{s}=(\phi_{1k}^{s}, \ldots,\phi_{Rk}^{s} )^{\top}$, denotes the group-specific factor loadings. 

The low-rank representation in \eqref{multi-group model}, with both shared and group-specific latent factors, enables efficient estimation while capturing common temporal patterns and group-specific deviations. We specify Gaussian independent priors on the latent factors as 
\begin{equation}
\begin{split}
    \boldsymbol{\eta}_{is} {\sim} \mathcal{N}_{L}(0_{L},\boldsymbol{I}_{L}),\\ 
   \boldsymbol{\rho}_{is} {\sim} \mathcal{N}_{K_{s}}(0_{K_{s}},\boldsymbol{I}_{K_{s}}).
\end{split}
\label{priors}
\end{equation}
Following \cite{Shamshoianetal2022}, we set the number of spline bases to $R \approx T/2$. Even with this choice, overfitting may occur if the basis is too flexible, especially for relatively smooth functional data. To avoid this, we adopt Penalized B-splines \citep{EilersandMarx1996},  which impose a roughness penalty on first- or second-order differences of adjacent coefficients. In our Bayesian formulation, this is implemented by assigning a Gaussian prior to the group-specific mean coefficients $\boldsymbol{\beta}_{s}$, as 
\begin{equation}
\label{eq:priorB}
\boldsymbol{\beta}_{s} | \sigma_{\beta_{s}}^{2} \sim \mathcal{N}(0,\sigma_{\beta_{s}}^{2}\Omega^{-1}),
\end{equation}
 where $\sigma_{\beta_{s}}^{2} \sim \text{Inv-Gamma}\left(a_{\beta_{s}}, b_{\beta_{s}}\right)$, and $\Omega$ is a penalty matrix that encodes smoothness, defined as $\Omega= \Omega^{*} + \varsigma \boldsymbol{I}_{R}$. Here, $\Omega^{*}$ represents a discrete roughness penalty based on second-order differences of adjacent spline coefficients, which is rank-deficient by construction \citep{Bolandetal2023}. The coefficient $\varsigma$ is a small constant (e.g. $\varsigma=0.0000001$) added to guarantee positive definiteness. The corresponding prior density is proportional to 
$\exp\left(-\frac{1}{2\sigma_{\boldsymbol{\beta}_{s}}^{2}} \boldsymbol{\beta}_{s}^{\top} \Omega \boldsymbol{\beta}_{s}\right)
$, for $s=1,\ldots, S$.

As a result of the model assumptions, the marginal distribution of the observed functional data $\boldsymbol{y}_{is}$ is multivariate normal, with mean $B \boldsymbol{\beta}_{s}$ and covariance 
$\Sigma_{Y_{s}} = \Sigma_{{\Lambda}} + \Sigma_{{\Phi}_{s}} + \Sigma_{\epsilon_{s}}$, with
\begin{equation}
    \label{eqSlam}
\Sigma_{{\Lambda}} = B\Lambda \Lambda^{\top}B^{\top},
\end{equation}
representing the shared covariance, which captures variability common to all groups, and 
\begin{equation}
    \label{eqSphi}
    \Sigma_{{\Phi}_{s}}=B\Phi_{s}\Phi_{s}^{\top}B^{\top},
\end{equation}
denoting the group-specific covariance, which captures variability unique to group $s$. The residual covariance is given by $\Sigma_{\epsilon_{s}}=\sigma_{\epsilon_{s}}^{2}\boldsymbol{I}_{T}$, where $\sigma_{\epsilon_{s}}^{2}$ is assigned a noninformative prior proportional to a constant $c$ \citep{Bolandetal2023}.

\subsection{Cumulative Shrinkage Process Priors for Shared and Group-Specific Factor Loadings}
\label{prior}
 
For the shared factor loading matrix, $\Lambda$, and the group-specific factor loading matrices, $\Phi_{s}$, we build on the parameter-expansion strategy of \citet{Scheipletal2012} to define a cumulative shrinkage process prior \citep[CUSP,][]{Legramantietal2020} that encourages adaptive shrinkage and allows the model to automatically select the number of active factors. By introducing auxiliary parameters, the proposed parameter-expanded cumulative shrinkage process prior better handles posterior dependences and improves mixing and convergence of MCMC samplers. A similar expansion approach is employed in \cite{KowalandCanale2023}, though in a single-population setup and as a prior on the factors, so that the shrinkage regularizes the distribution of subject-scores instead of the latent trajectories that define the functional structure. 

We start by considering the shared factor loading matrix $\Lambda$ and the parameterization,
\begin{equation}
\label{eq7}
\Lambda =\Xi \times \text{diag}(\gamma_{1},\ldots, \gamma_{L}),
\end{equation}
where $\Xi=(\boldsymbol{\xi}_{1},\ldots,\boldsymbol{\xi}_{L})$ is an $R\times L$ auxiliary matrix with entries $\xi_{rl}$, and $(\gamma_{1}, \ldots,\gamma_L)$ is a vector of global scale parameters controlling the relevance of each of the $L$ factors and, therefore, each latent functional component. The entries of $\Xi$ capture the relative contributions of the basis functions that define each latent trajectory $\widetilde{\lambda}_{l}(t)$, effectively shaping the form of each trajectory. They are modeled as 
\begin{equation}
    \xi_{rl}|m_{rl} \sim \mathcal{N}(m_{rl},1), \quad  m_{rl} \sim \frac{1}{2}\delta_{1}(m_{rl}) + \frac{1}{2}\delta_{-1}(m_{rl}),
    \label{eq_xi_prior}
\end{equation}
where $\delta_{1}(\cdot)$ and $\delta_{-1}(\cdot)$ denote point masses at 1 and -1, respectively. Compared to the standard choice of $\xi_{rl} \sim \mathcal{N}(0,1)$, this leads to a marginal prior for $\gamma_{l}$ still centered at zero but that places less weight on values close to zero, which improves MCMC mixing \citep{KowalandCanale2023}. For each global parameter $\gamma_l$, we specify a cumulative shrinkage prior that progressively shrinks higher-order columns in $\Lambda$ toward zero, ensuring that redundant factors as well as functional components are pruned away while maintaining flexibility in the active components. More specifically, we assume
\begin{equation}
    \gamma_{l} \sim \mathcal{N}(0, \theta_{l}\sigma_{\gamma_l}^{2}), \quad \theta_{l}|\pi_{l} \sim (1-\pi_{l})\delta_{1} + \pi_{l}\delta_{v_{0}}, \quad \sigma_{\gamma_l}^{2} \sim \text{Inv-Gamma}(a_{1},a_{2}),
\label{eq_gamma_prior}
\end{equation}
for $l=1,\ldots,L$, with $\delta_{1}(\cdot)$ and $\delta_{v_0}(\cdot)$ denoting point mass at 1 and $v_0$, respectively. Here,  $\theta_{l}$ acts as a variance component that determines whether the $l$-th column of $\Lambda$ is active or effectively suppressed: when $\theta_{l}=1$, the variance is governed by $\sigma_{\gamma_l}^{2}$; 
when $\theta_{l}=v_{0}\ll 1$, the variance is strongly reduced, shrinking the column of the loading matrix toward zero. The additional variance parameter $\sigma_{\gamma_l}^{2}$ introduces further adaptability for active factor loadings by allowing their magnitudes to vary with $l$, while a very small value of $v_{0}$ ensures that inactive columns are heavily constrained toward zero.  Marginalizing over $\theta_{l}$ and $\sigma_{\gamma_l}^{2}$, conditional on $\pi_l$, leads to a bimodal mixture of scaled $t$-distributions
\begin{equation}
[\gamma_{l}|\pi_{l}] \sim (1-\pi_{l})\, t_{2a_{1}}\left(0, \sqrt{a_{2}/a_{1}}\right) + \pi_{l}\, t_{2a_{1}}\left(0, \sqrt{v_{0}a_{2}/a_{1}}\right),
\end{equation}
with degrees of freedom  $df=2a_{1}$, and scale parameters $a_{2} \gg a_{1}$,  chosen to induce strong shrinkage on inactive factors, by virtue of negligible factor loadings. This prior structure is inspired by the normal mixture of inverse gamma priors of \cite{IshwaranandRao2005}. Furthermore, the prior encourages the columns of the loading matrix to shrink toward zero with increasing probability as $l$ increases, $l=1, \ldots, L$, allowing the model to adaptively identify the number of active factors.
More specifically, the column-specific shrinkage probabilities $\pi_{l}$ are generated via a stick-breaking construction,
\begin{equation}
\pi_{l}=\sum_{h=1}^{l}\omega_{h}, \quad 
\omega_{h}=\nu_{h}\prod_{m=1}^{h-1}(1-\nu_{m}), \quad l=1, \ldots, L,
\label{eq_stick_breaking_prior}
\end{equation}
with $\nu_{h} \stackrel{\textit{iid}}{\sim} \text{Beta}(\iota,\iota \alpha)$, and $\alpha \sim \text{Gamma}(a_{\alpha}, b_{\alpha})$,
which naturally assigns decreasing prior probabilities of activation to higher indexes, hereby promoting shrinkage of redundant columns in $\Lambda$. 
We fix $\iota=1$, following \cite{KowalandCanale2023}.  The weights $\omega_{l}$ satisfy $\sum_{h=1}^{L} \omega_{h} = 1$. In practice, we truncate the construction by fixing $\nu_{L}=1$ for some finite $L < \infty$, which yields a finite representation of the CUSP prior.
 
For computational purposes, it is convenient to introduce a latent indicator $z_{l}$ for each factor $l$, drawn from a categorical distribution with weights $\{\omega_{h}\}$, such that $\Pr(z_{l}=h)=\omega_{h}$. Conditional on $z_{l}$, the variance component $\theta_{l}$ is specified as
\begin{equation*}
[\theta_{l} | z_{l}] \sim \left(1-\mathbb{I}(z_{l} \le l)\right)\, \delta_{1} + \mathbb{I}\left(z_{l} \le l\right)\, \delta_{v_0}.
\end{equation*}
The number of active shared factors can then be estimated by evaluating 
$L^{*} = \sum_{l=1}^{\infty}\mathbb{I}(z_{l}>l)$.

We impose a similar structure on the group-specific factor loading matrices $\Phi_{s}$, for each group $s$, as
\begin{equation}
\Phi_{s} =\Xi^{s} \times \text{diag}(\gamma^{s}_{1},\ldots, \gamma^{s}_{K_{s}}), 
\label{eq8}
\end{equation}
where $\Xi^{s}=(\boldsymbol{\xi}^{s}_{1},\ldots,\boldsymbol{\xi}^{s}_{K_s})$ is an $R\times K_s$ auxiliary matrix with entries $\xi^{s}_{rk}$.
We assign the priors
\begin{gather}
\xi^{s}_{rk}|m^{s}_{rk} \sim \mathcal{N}(m^{s}_{rk},1), \quad m^{s}_{rk} \sim \frac{1}{2}\delta_{1}(m^{s}_{rk}) + \frac{1}{2}\delta_{-1}(m^{s}_{rk}),
\label{eq_xi_s_prior}\\
\gamma^{s}_{k} \sim \mathcal{N}(0, \theta^{s}_{k}\sigma_{\gamma^{s}_{k}}^{2}), \quad \theta^{s}_{k}|\pi^{s}_{k} \sim (1-\pi^{s}_{k})\delta_{1} + \pi^{s}_{k}\delta_{v^{s}_{0}}, \quad \sigma_{\gamma^{s}_{k}}^{2} \sim \text{Inv-Gamma}(a^{s}_{1},a^{s}_{2}), 
\label{eq_gamma_s_prior}\\
\pi^{s}_{k}=\sum_{h=1}^{k}\omega^{s}_{h}, \quad
 \omega^{s}_{h}=\nu^{s}_{h}\prod_{m=1}^{h-1}(1-\nu^{s}_{m}), \quad \nu^{s}_{m} \stackrel{\textit{iid}}{\sim} \text{Beta}(\iota^{s}, \iota^{s} \alpha^{s}), \quad \alpha^{s} \sim \text{Gamma}(a^{s}_{\alpha}, b^{s}_{\alpha}),
 \label{eq_stick_breaking_s_prior}
\end{gather}
and estimate the number of active group-specific factors by evaluating
$K_{s}^{*} = \sum_{k=1}^{\infty}\mathbb{I}(z_{k}^{s}>k)$ for $s=1,\ldots,S$, with $z_{k}^{s}$ latent indicators drawn from categorical distributions with weights $\omega^s_{k}$. 

As pointed out in \cite{Scheipletal2012}, parameters $\gamma_{l}$ and $\xi_{rl}$ in equation~\eqref{eq7} and parameters $\gamma^s_{k}$ and $\xi^s_{rk}$ in equation~\eqref{eq8} are not identifiable, indicating that their individual values can drift to extreme regions of the parameter space without impacting the overall model fit. We follow their suggestion and jointly rescale these parameters at each MCMC iteration. 

\subsection{Markov chain Monte Carlo Algorithm}

We employ a Markov chain Monte Carlo (MCMC) algorithm to draw samples from the posterior distribution of the model parameters. A schematic description of the algorithm is provided as Algorithm 1, while full details are given in the Supplementary Materials. 

\begin{algorithm}
{\footnotesize
\label{alg:mcmc}
\SetAlgoLined
\LinesNotNumbered 
\SetKwInOut{Input}{Input}
\SetKwInOut{Output}{Output}
\Input{Data $Y_{s}$, for $s=1,\ldots, S$ and $i=1,\ldots,n_{s}$.}

\Output{Posterior samples for $\boldsymbol{\beta}_{s}, \sigma^{2}_{\epsilon_{s}}, \sigma^{2}_{\beta_{s}}, \Lambda, \boldsymbol{\eta}_{is}, \sigma_{\gamma}^{2}, z, \nu, \theta, \alpha, \Phi_{s}, \boldsymbol{\rho}_{is}, \sigma_{\gamma^{s}}^{2}, z^{s}, \nu^{s}, \theta^{s}, \alpha^{s}$} 

\textbf{Sample} $\boldsymbol{\beta}_{s}$ from normal full conditional distribution. 

\textbf{Sample} $\sigma_{\epsilon_{s}}^{2}$ from inverse-gamma full conditional distribution. 

\textbf{Sample} $\sigma_{\beta_{s}}^{2}$ from inverse-gamma full conditional distribution.

\textbf{For shared components}:\\
\Indp
  \Begin{
    \textbf{Sample} $m$ from the discrete uniform on $\{-1,1\}$.
        
    \textbf{Sample} $\Xi$ from the multivariate normal full conditional distribution.
    
\For{$l=1$ \KwTo $L$}{        
    \textbf{Sample} $\gamma_{l}$ from the normal full conditional distribution, given the most recently updated values of $\gamma_{1}, \ldots, \gamma_{l-1}$ and the previous iteration's values of $\gamma_{l+1}, \ldots, \gamma_{L}$.
}

\For{$l=1$ \KwTo $L$}{  
\textbf{Rescale} $\boldsymbol{\xi}_{l}$
and $\gamma_{l}$.}

\textbf{Update} $\Lambda=\Xi \cdot \text{diag}(\gamma_{1}, \ldots, \gamma_{L})$.

\textbf{Sample} $\boldsymbol{\eta}_{is}$ from the  normal full conditional distribution.

\For{$l=1$ \KwTo $L$}{
\textbf{Sample} $\sigma_{\gamma_{l}}^{2}$ from the inverse-gamma full conditional distribution.

\textbf{Sample} $ z_{l}, \nu_{l}, \theta_{l}$ from the DP stick-breaking construction. 
}

\textbf{Sample} $\alpha$ from the gamma full conditional distribution.
} 
\Indm 

\textbf{For group-specific components}:\\
\Indp
  \Begin{
    \textbf{Sample} $m^{s}$ from the discrete uniform on $\{-1,1\}$.
        
    \textbf{Sample} $\Xi^{s}$ from the multivariate normal full conditional distribution.
    
    \For{$k=1$ \KwTo $K_s$}{    
        \textbf{Sample} $\gamma_{k}^{s}$ from the normal full conditional distribution, given the most recently updated values of $\gamma^{s}_{1}, \ldots, \gamma^{s}_{k-1}$ and the previous iteration's values of $\gamma^{s}_{k+1}, \ldots, \gamma^{s}_{K_{s}}$.
    }   
    \For{$k=1$ \KwTo $K_s$}{ 
    \textbf{Rescale} $\boldsymbol{\xi}_{k}^{s}$ and $\gamma_{k}^{s}$.
    }
    \textbf{Update} $\Phi_{s}=\Xi^{s} \cdot \text{diag}(\gamma^{s}_{1}, \ldots, \gamma_{K_s}^{s})$.
        
    \textbf{Sample} $\boldsymbol{\rho}_{is}$ from the normal full conditional distribution.
        
    \For{$k=1$ \KwTo $K_s$}{   
        \textbf{Sample} $\sigma_{\gamma^{s}_{k}}^{2}$ from the inverse-gamma full conditional distribution.
        
        \textbf{Sample} $z_{k}^{s}, \nu_{k}^{s}, \theta_{k}^{s}$ from the DP stick-breaking construction.
    }
    \textbf{Sample} $\alpha^{s}$ from the gamma full conditional distribution.
}
\Indm
\caption{Gibbs sampler for BMGFFM.}
}
\end{algorithm}

\subsection{Posterior Inference}
\label{Posterior Inference}
Posterior inference for latent factor models is complicated by well-known non-identifiability issues, including invariance to orthogonal rotations, sign switching, and permutations of the factor loadings. In addition, in the present multi-group setting, the decomposition of the latent covariance into shared and group-specific components is not uniquely identifiable at the level of the loading matrices, since different parameterizations of the shared and group-specific loadings can induce the same covariance structure \citep{DeVitoetal2021}.
To address these issues, we adopt a post-processing strategy that separates (i) alignment of posterior draws in parameter space from (ii) identification of interpretable loading structures in covariance space. 
This procedure leads to identifiable and interpretable estimates of the shared and group-specific loading structures by operating in covariance space rather than directly on the factor loadings. 

Given the MCMC output, we first examine the active factor configurations at each iteration and select the combination that occurs most frequently across iterations, after burn-in,  as the optimal number of active factors, $[L^{*}, K_{1}^{*}, \ldots, K^{*}_{s}]$. We then retain only those MCMC iterations where the number of active factors matches this optimal factor structure for posterior inference.  
Next, we apply a post-processing strategy to address identifiability issues, such as rotational invariance, sign switching, and columns permutation (also known as label-switching), which are common issues in factor models. Rotational invariance refers to the fact that, for any orthogonal matrix, the transformed factor loadings and transformed factor scores yield the same marginal likelihood. Sign switching occurs because changing the sign of a column in the factor loadings matrix and of the corresponding factor score simultaneously leaves the likelihood unchanged. Permutation invariance refers to the fact that the ordering of the latent factors is arbitrary. Because of these issues, averaging MCMC samples of the factor loadings and scores matrices without post-processing of the samples may produce distorted or uninterpretable estimates. Several post-processing approaches have been proposed \citep{Stephens2000, Fruhwirth2011dealing, Poworozneketal2021}. Here we apply the rotation sign permutation (RSP) algorithm proposed by \cite{PapastamoulisandNtzoufras2022}, which gave us the most consistent results across MCMC chains. This method iteratively performs rotation, sign, and column permutation across posterior samples. 
First, each sampled factor loading matrix is rotated using the \texttt{varimax()} function \citep{Kaiser1958varimax}, by multiplying it with an orthogonal rotation matrix that maximizes the within-factor variance of squared loadings. The rotated matrices are then iteratively aligned across MCMC draws via a signed permutation procedure that minimizes the element-wise discrepancy to a reference loading matrix.  The corresponding factor scores are also relabeled by applying the same varimax rotation, sign, and permutation matrices. This step is repeated until convergence. The algorithm is implemented in the \texttt{factor.switching} package in \texttt{R}.
In our implementation, we apply the RSP algorithm in the coefficient space to the shared loadings and scores, $\Lambda^{(m)}$ and $\boldsymbol{\eta}_{is}^{(m)}$, as well as the group-specific components, $\Phi_s^{(m)}$ and $\boldsymbol{\rho}_{is}^{(m)}$, within each group $s$, across the retained posterior draws.  Estimated latent curves are obtained by averaging over the posterior relabeled samples as $\widehat{\boldsymbol{f}}_{is} = \frac{1}{|\mathcal{M}|} \sum_{m \in \mathcal{M}} B\Big(\boldsymbol{\beta}_s^{(m)} + \Lambda^{(m)} \boldsymbol{\eta}_{is}^{(m)} + \Phi_s^{(m)} \boldsymbol{\rho}_{is}^{(m)}\Big)$, with $\mathcal{M}$ the set of posterior draws with the modal factor configuration. These estimates are invariant to the choice of loading parameterization.

Finally, inspired by \cite{DeVitoetal2021}, rather than interpreting the loading matrices directly, we summarize the shared and group-specific structures through their induced covariance operators.
Specifically, we compute posterior mean covariance operators for the shared and group-specific components as
\begin{eqnarray}
\widehat{\Sigma}_\Lambda =  \frac{1}{|\mathcal{M}|} \sum_{m \in \mathcal{M}}  B \Lambda^{(m)} \Lambda^{(m)\top} B^\top,
\quad
\widehat{\Sigma}_{\Phi_s} = \frac{1}{|\mathcal{M}|} \sum_{m \in \mathcal{M}}  B \Phi_s^{(m)} \Phi_s^{(m)\top} B^\top,
\end{eqnarray}
and define the total latent covariance for group \(s\) as $\widehat{\Sigma}_{f_s} = \widehat{\Sigma}_\Lambda + \widehat{\Sigma}_{\Phi_s}$.
We then obtain identifiable and interpretable latent subspaces, together with associated covariance-derived loadings, via the spectral decomposition $\widehat{\Sigma}_\Lambda = U_\Lambda D_\Lambda U_\Lambda^\top$, where $U_\Lambda$ is the matrix of eigenvectors and $D_\Lambda$ is the diagonal matrix of corresponding eigenvalues. Let $U_{\Lambda,L^*}$ and $D_{\Lambda,L^*}$ denote the leading $L^*$ eigenvectors and eigenvalues of $\widehat{\Sigma}_\Lambda$, respectively. The shared latent subspace is then defined as
$\widehat{\mathcal S}_\Lambda = \mathrm{span}(U_{\Lambda,L^*})$, and the corresponding covariance-derived shared loadings are defined by
$\widetilde{\Lambda}_{L^*} = U_{\Lambda,L^*} D_{\Lambda,L^*}^{1/2}$.
For each group $s$, the group-specific latent structure is obtained from the residual covariance
$\widehat{\Sigma}_{\mathrm{res},s} = \widehat{\Sigma}_{f_s} - \widetilde{\Lambda}_{L^*} \widetilde{\Lambda}_{L^*}^{\top}$,
with spectral decomposition $\widehat{\Sigma}_{\mathrm{res},s} = U_s D_s U_s^\top$, where $U_s$ is the matrix of eigenvectors and $D_s$ is the diagonal matrix of corresponding eigenvalues. Let $U_{s,K_s^*}$ and $D_{s,K_s^*}$ denote the leading $K_s^*$ eigenvectors and eigenvalues of $\widehat{\Sigma}_{\mathrm{res},s}$, respectively. The group-specific latent subspace is defined as
$\widehat{\mathcal S}_{\Phi_s} = \mathrm{span}(U_{s,K_s^*})$, and the corresponding covariance-derived group-specific loadings are defined by $\widetilde{\Phi}_{s,K_s^*} = U_{s,K_s^*} D_{s,K_s^*}^{1/2}$.
This procedure yields an orthogonal decomposition of the latent covariance into shared and group-specific subspaces. Importantly, the resulting loading spaces are uniquely defined (up to sign) by the covariance operators, thereby avoiding the need for additional identification constraints or projection steps in parameter space.

%---------------------------------------------------
\section{Simulation Study}
\label{sec:Simulation}
We conduct a simulation study to evaluate the performance of the proposed model and compare it against alternative methods. 

\subsection{Data Generation}
\label{subsec:gatagen}
We set $S=2$ and consider three main scenarios to assess how performance varies with the number of shared ($L$) and group-specific ($K_s$) factors:
\begin{itemize}
\item \textbf{Scenario~A:} In this scenario we allow for both shared and group specific factors. We set $L=3$ shared factors and consider two cases for the number of group-specific factors, one with $K_1=K_2=2$, and one with $K_1=2, K_2=0$.

\item \textbf{Scenario~B:} This case considers a homogeneous latent structure across groups, with $L=3$ shared factors and no group-specific factors, i.e., $K_1=K_2=0$. 

\item \textbf{Scenario~C:} This is the opposite case to Scenario~B, with complete heterogeneity between groups, that is $L=0$ shared factors and $K_1=K_2=2$. 

\end{itemize}
In each scenario, we consider three cases: $(a)$ $n_1=40, n_2=40$, $(b)$ $n_1=80, n_2=80$, and $(c)$  $n_1=40, n_2=80$. We set $T = 60$ and $R=T/2=30$ and generate the functional curves from model \eqref{multi-group model} as follows. First, we generate the group-specific coefficients $\boldsymbol{\beta}_{s}$ from $\mathcal{N}(0,\sigma_{\beta_{s}}^{2})$, for $s=1,2$,  with variances $\sigma_{\beta_{1}}^{2}=0.2$, and $\sigma_{\beta_{2}}^{2}=0.4$, and generate $\Lambda$ and $\Phi_{s}$ from multivariate standard normal distributions. Then, we sample the latent signals $\boldsymbol{f}_{is}$ once from the distribution 
\begin{equation*}
\mathcal{N}(B\boldsymbol{\beta}_{s}, \Sigma_{f_{s}}), ~~\Sigma_{f_{s}} = B\Lambda\Lambda^{\top}B^{\top} + B\Phi_{s} \Phi_{s}^{\top}B^{\top},~~ s=1,2,
\end{equation*}
and keep them fixed across 100 replicated datasets, which we generate by adding independent noise drawn as $\boldsymbol{\epsilon}_{is} \sim \mathcal{N}(0, \sigma_{\epsilon_{s}}^{2}\boldsymbol{I}_{T})$, with $\sigma_{\epsilon_{s}}^{2}$ set to achieve a pre-fixed signal-to-noise ratio (SNR), defined as the ratio between the variance explained by the latent signals and the residual variance as $\text{SNR}=\text{tr}(\Sigma_{f_{s}})/(T\sigma_{\epsilon_{s}}^{2})$, in each group \citep{Nguyenetal2024}. Below we provide results for SNR=2 and report full results for SNR=5 in the Supplementary Materials.

\subsection{Parameter Settings}

When applying our model, we set the prior hyperparameters as follows. For the group-specific mean coefficients $\boldsymbol{\beta}_{s}$, we set the parameters of the hyperprior on $\sigma_{\beta_{s}}^{2}$ as $a_{\beta_{s}}=b_{\beta_{s}}=1$, following \cite{Bolandetal2023}. 
For the expected number of shared and group-specific factors, we set the hyperparameters pairs $(a_{\alpha}, b_{\alpha})$ in equation \eqref{eq_stick_breaking_prior} and $(a_{\alpha}^{s}, b_{\alpha}^{s})$ in equation \eqref{eq_stick_breaking_s_prior} to $(2,1)$, following  \cite{KowalandCanale2023}. For the priors on the shared and group-specific factor loading matrices, we set the hyperparameters $(a_{1}, a_{2}, v_{0})$ in equation \eqref{eq_gamma_prior} and $(a_{1}^{s}, a_{2}^{s}, v_{0}^{s})$ in equation \eqref{eq_gamma_s_prior} to $(10,30,0.001)$. We discuss results from a sensitivity analysis on these choices in Section \ref{sensitivity analysis}. 

MCMC chains were run for $20,000$ iterations, with the first $10,000$ iterations discarded as burn-in. 
We initialized the sampler with an upper bound of $L_{\max}=10$ shared factors and $K_{s,\max}=10$ group-specific factors for each group, allowing the shrinkage process prior to learn the numbers of active factors by shrinking inactive factors toward zero. 
To assess convergence, we monitored model parameters using Geweke's convergence diagnostic \citep{Geweke1992evaluating} and by visually inspecting traceplots. For example, we examined the parameters $\sigma^{2}_{\epsilon_{s}}$ and $\sigma^{2}_{\beta_{s}}$ for randomly selected replicates from Scenario~A~$(3,2,2)$ with $n_1 = 40$ and $n_2 = 80$. These parameters showed Geweke diagnostics well below the critical threshold of $\pm 1.96$. Additionally, traceplots for randomly selected parameters revealed no significant drift and demonstrated good mixing across iterations for the selected parameters (see Supplementary Materials). Furthermore, inspection of representative elements of $\boldsymbol{\beta}_{s}$ revealed acceptable diagnostics and stable behavior (see Supplementary Materials). Overall, these diagnostics indicate that the MCMC sampler achieved adequate convergence across key components of the model, providing reliable posterior inference.

\subsection{Results}

\begin{figure}
    \centering
    \includegraphics[width=1\linewidth]{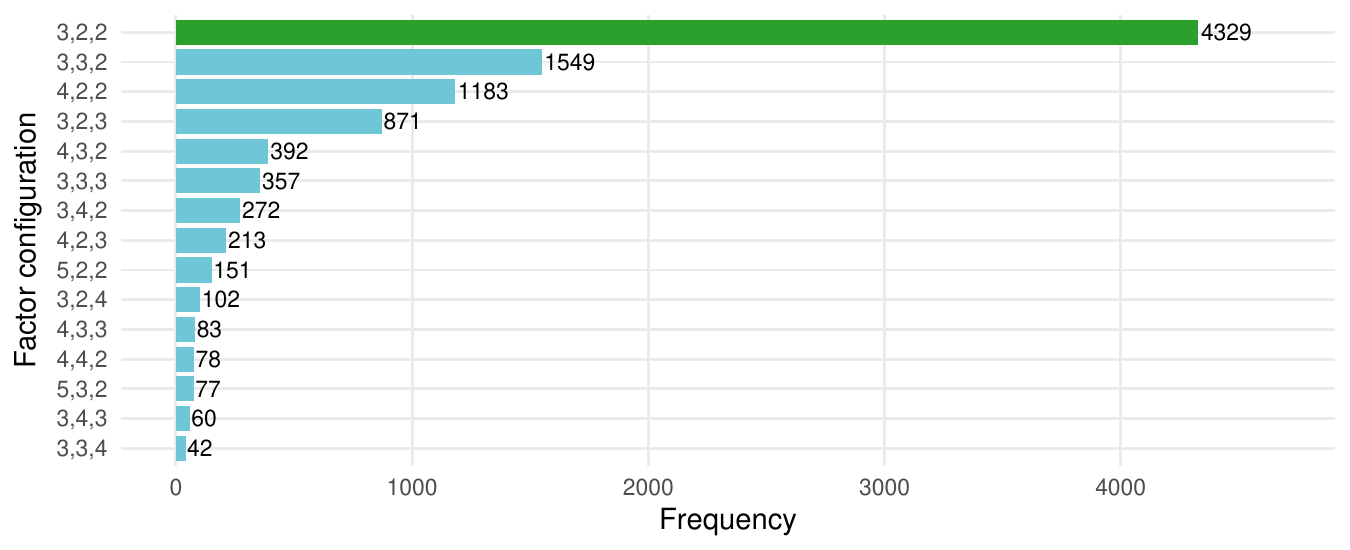}
    \caption{\textbf{Simulation Study.} Posterior frequencies of the $15$ most frequently sampled factor configurations for one replicate from Scenario~A~$(3,2,2)$ with $n_{1}=40$, and $n_{2}=80$, based on $10,000$ retained MCMC draws after burn-in. The modal factor configuration coincides with the true configuration $(3,2,2)$.}
    \label{fig:Figure2_fac_conf_sceA3}
    \end{figure}

\begin{figure}
    \centering
    \includegraphics[width=0.95\linewidth]{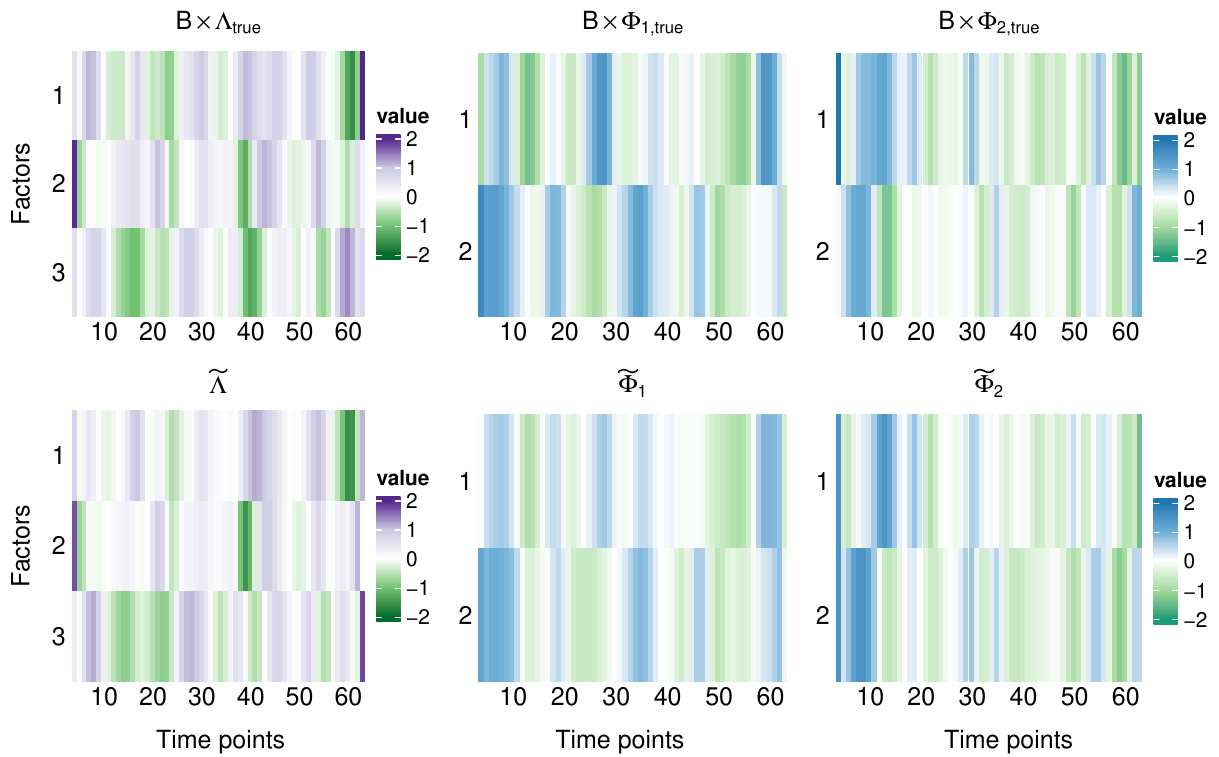}
    \caption{\textbf{Simulation Study}. Comparison of heatmaps for the true (top) and estimated (bottom) covariance-derived shared and group-specific factor loadings for the same replicate from Scenario~A~$(3,2,2)$ with $n_{1}=40$, and $n_{2}=80$ shown in Figure~\ref{fig:Figure2_fac_conf_sceA3}.}
    \label{fig:SNR2_simu_heatmaps_loadings}
\end{figure}

\begin{figure}
        \centering
        \includegraphics[width=0.9\linewidth]{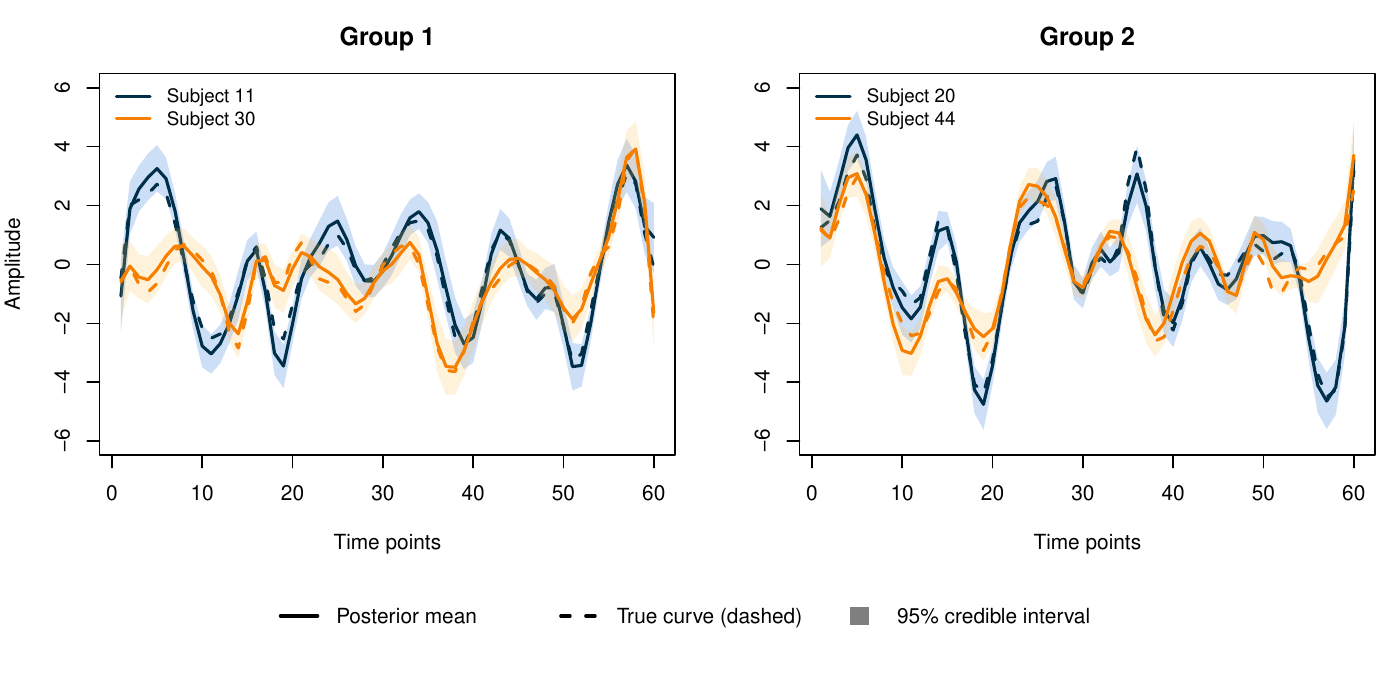}
        \caption{\textbf{Simulation Study.} True and estimated curves $f_{s}$, together with pointwise $95\%$ credible intervals, for two representative subjects from each group for the same replicate from Scenario~A~$(3,2,2)$ with $n_{1}=40$, and $n_{2}=80$ shown in Figure~\ref{fig:Figure2_fac_conf_sceA3}.}
    \label{fig:sceA3_SNR2_rep76_representative_subjects_two_panels}
    \end{figure}

\begin{figure}
    \centering
    \includegraphics[width=1\linewidth]{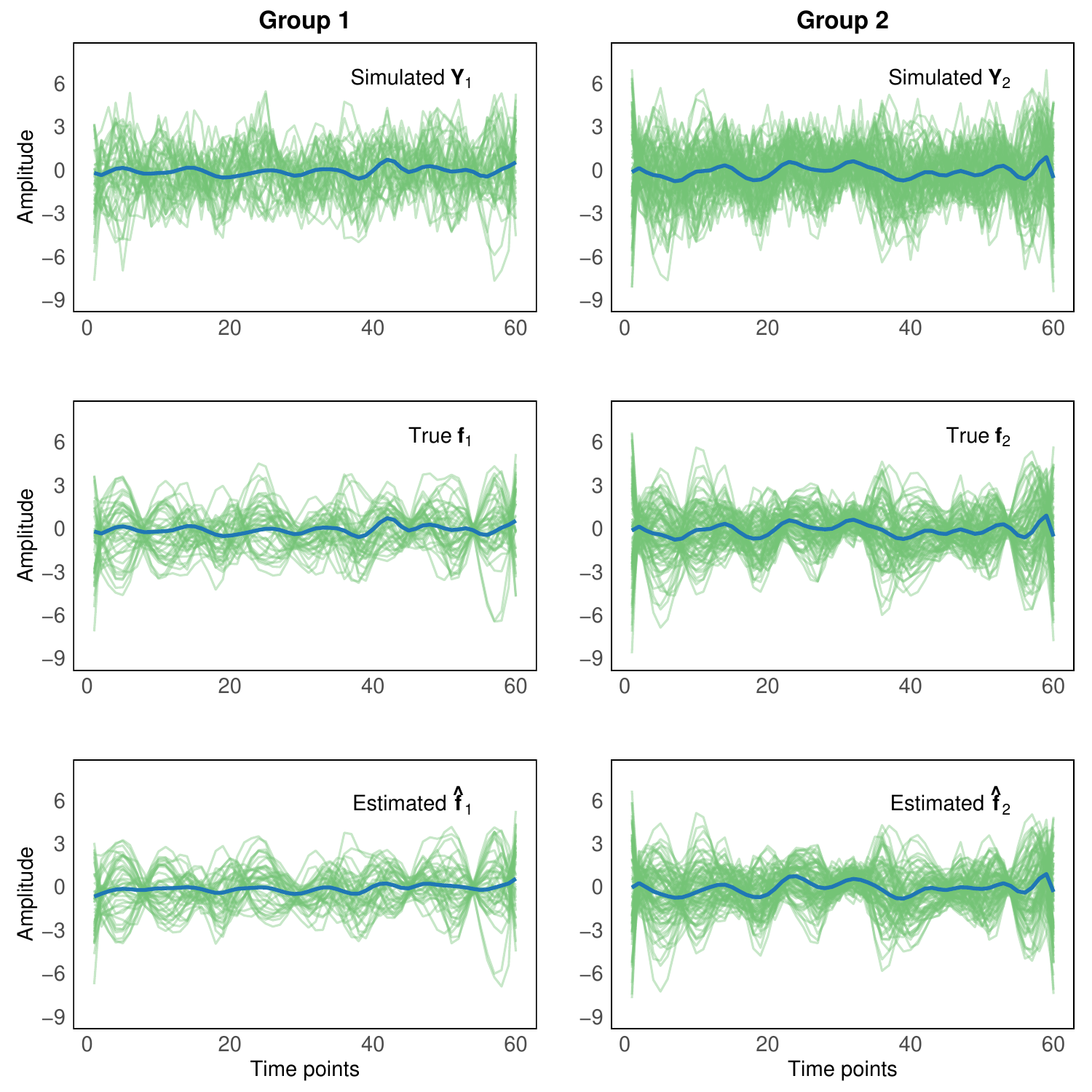}
    \caption{\textbf{Simulation Study}. Comparison of simulated curves, $Y_s$, and true and estimated curves $f_{s}$ for all subjects in the two groups for the same replicate from Scenario~A~$(3,2,2)$ with $n_{1}=40$, and $n_{2}=80$ shown in Figure~\ref{fig:Figure2_fac_conf_sceA3}. Top line: simulated $Y_s$ for 40 subjects (left) and 80 subjects (right) over the 60 time points, with subject-level curves in green and averaged curves in blue; Middle line: true curves $f_{s}$ for the two groups, with subject-level curves in green and averaged curves in blue; Bottom line: estimated curves $\widehat{f}_{s}$ , shown as posterior means, with subject-level curves in green and averaged curves in blue.} 
\label{fig:SNR2_simu_true_estimated_curves}
\end{figure}

We start by showing results for one replicated dataset from Scenario~A (3,2,2) with $n_1=40$, $n_2=80$.  Figure \ref{fig:Figure2_fac_conf_sceA3} reports the posterior distribution of the factor configurations across $10,000$ retained MCMC draws after burn-in, restricted to the 15 most frequently sampled factor configurations. For this dataset, our method correctly estimates the number of latent factors as $L=3$ and $K_1=K_2=2$. For the same replicated dataset, Figure~\ref{fig:SNR2_simu_heatmaps_loadings} shows heatmaps of the true loading matrices and the estimated ones, illustrating that the model provides close recovery of both the shared and group-specific structures.  In Figure~\ref{fig:sceA3_SNR2_rep76_representative_subjects_two_panels} we report true and estimated curves, together with pointwise $95\%$ credible intervals, for two representative subjects from each group, while Figure~\ref{fig:SNR2_simu_true_estimated_curves} shows simulated, true and estimated curves for all subjects in the two groups, together with the mean curves by group, indicating that posterior estimates closely approximate the underlying signals. 

Next, we comment on the results from the estimation of the numbers of shared and group-specific factors, $L$ and $K_s$ for $s=1,2$, averaged across 100 replicates, for all scenarios. These results are summarized in Table~\ref{tab:SNR2_esti_num_factors}. Overall, the posterior estimates are close to the true values, and the results show consistent estimation performance across different sample size settings. For scenario~A, where both shared and group-specific factors are present, the model generally recovers the correct dimensionality, with some moderate variability across replicates. In Scenario~B, which considers a homogeneous latent structure, the model accurately identifies the number of shared factors and correctly estimates the absence of group-specific factors in most replicates. Conversely, Scenario~C, which represents complete heterogeneity between groups, is the most challenging. In this scenario, our model still correctly detects the absence of shared structure, while recovering the group-specific factors, but with larger variability across replicates with respect to the previous scenarios.  This scenario is particularly challenging, since no shared structure is present. In this case, the model must infer the absence of common factors while allocating all systematic variation to the group-specific components, which makes factor number estimation less stable than in the other scenarios. This difficulty is more pronounced in the unbalanced $\{40,80\}$ setting, where the asymmetry in sample sizes further increases estimation uncertainty.

\begin{table}[!htbp]
    \centering
    \caption{\textbf{Simulation Study}. Estimated number of shared factors ($L^{*}$) and group-specific factors, ($K^{*}_{s}$), reported as means and standard deviations, in parentheses, across 100 replicated datasets, for all simulated scenarios described in Section \ref{subsec:gatagen}. }
    \begin{tabular}{c c c c c c }
    \Xhline{1pt}
        \textbf{Scenario} & ($L$, $K_1$, $K_2$) & $\{n_1,n_2\}$ & $L^{*}$ & $K_1^{*}$ & $K_2^{*}$  \\ 
        \addlinespace[1mm]
        \hline
        \addlinespace[1mm]
        A & (3,2,2) & \{40,40\}  & 2.82 (0.39) & 2.20 (0.40) & 2.00 (0.28) \\ 
        A & (3,2,2) & \{80,80\}   & 2.75 (0.50) & 2.03 (0.41) & 1.98 (0.40) \\ 
        A & (3,2,2) & \{40,80\}   & 3.13 (0.66) & 2.16 (0.37) & 1.70 (0.61) \\ 
        A & (3,2,0) & \{40,40\}   & 2.75 (0.46) & 2.27 (0.51) & 0.07 (0.26) \\ 
        A & (3,2,0) & \{80,80\} & 2.72 (0.51) & 2.01 (0.52) & 0.09 (0.32) \\ 
        A & (3,2,0) & \{40,80\}  & 2.65 (0.52) & 2.28 (0.51) & 0.10 (0.30) \\ 
        \addlinespace[1mm]
        \hline
        \addlinespace[1mm]
        B & (3,0,0) & \{40,40\}   & 2.71 (0.48) & 0.04 (0.20) & 0.04 (0.20) \\ 
        B & (3,0,0) & \{80,80\}   & 2.79 (0.41) & 0.02 (0.14) & 0.02 (0.14) \\ 
        B & (3,0,0) & \{40,80\}  & 2.81 (0.39) & 0.03 (0.17) & 0.03 (0.17) \\ 
        \addlinespace[1mm]
        \hline
        \addlinespace[1mm]
        C & (0,2,2) & \{40,40\}   & 0.07 (0.26) & 1.82 (0.39) & 1.83 (0.38) \\ 
        C & (0,2,2) & \{80,80\}  & 0.54 (0.72) & 1.58 (0.57) & 1.60 (0.57) \\
        C & (0,2,2) & \{40,80\}   & 0.73 (0.64) & 1.85 (0.30) & 1.14 (0.65) \\ 
    \Xhline{1pt} 
    \end{tabular}
    \label{tab:SNR2_esti_num_factors}
\end{table}

\begin{figure}[ht]
    \centering
    \includegraphics[width=0.8\linewidth]{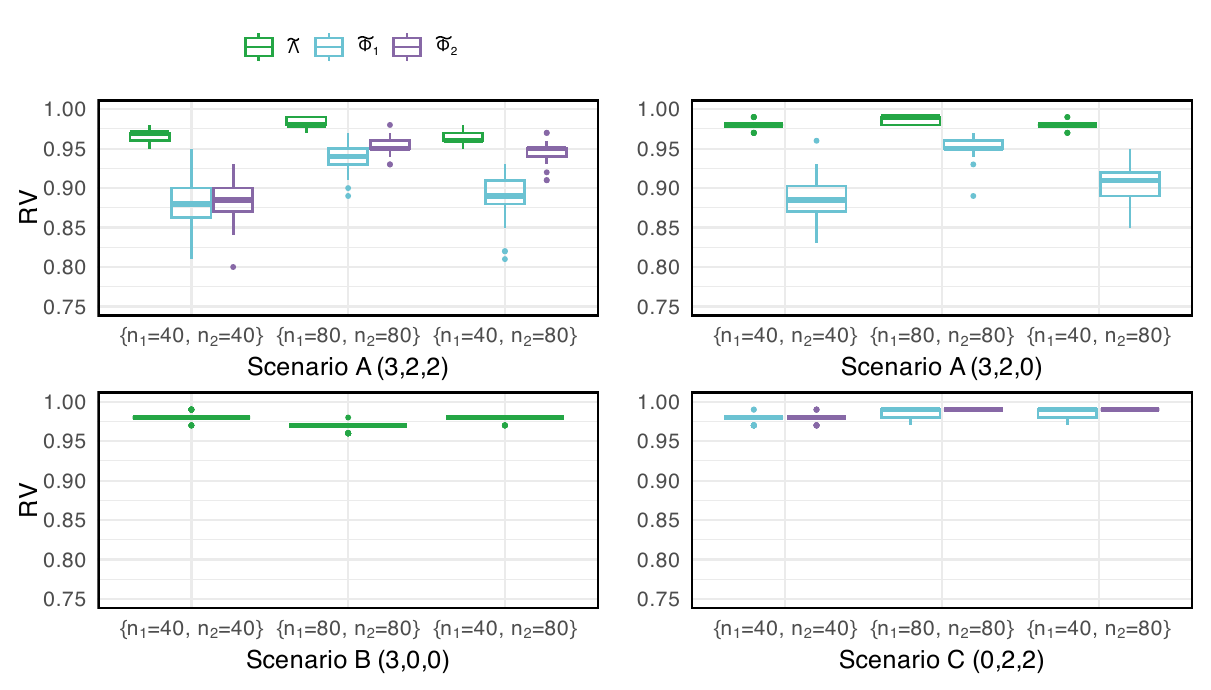}
    \caption{\textbf{Simulation Study}. Boxplots of the RV coefficients between the true shared and group-specific factor loadings, $B\Lambda_{\mathrm{true}}$ and $B\Phi_{s,\mathrm{true}}$, and the estimated covariance-derived shared and group-specific factor loadings, $\widetilde{\Lambda}$ and $\widetilde{\Phi}_{s}$, averaged across replicates where the estimates of $L^{*}$, $K_{1}^{*}$, and $K_{2}^{*}$ equal the true values, among 100 simulated datasets, for all scenarios described in Section~\ref{subsec:gatagen}.}
    \label{fig:SNR2_Boxplots_loadings}
\end{figure}

Next, we look at the estimates of the covariance-derived shared and group-specific factor loadings, $\widetilde{\Lambda}$, $\widetilde{\Phi}_{1}$ and $\widetilde{\Phi}_{2}$. For this inference, in order to report estimates across replicated datasets, we condition upon a correct estimation of $L, K_1$ and $K_2$. We then assess similarity between true and estimated matrices using the RV coefficient \citep{RobertandEscoufier1976, Abdi2007rv}.  Since factor loading matrices are rectangular, we first transform them into positive semi-definite matrices by multiplying each with its transpose, i.e., $E=XX^{\top}$ and $T=YY^{\top}$. The RV coefficient between the estimated matrix $X$ and true matrix $Y$ is then computed as
\begin{equation*}
    \text{RV} = \frac{\text{trace}(E^{\top}T)}{\sqrt{(\text{trace}(E^{\top}E)) \times (\text{trace}(T^{\top}T))}} =\frac{\text{trace}(XX^{\top}YY^{\top})}{\sqrt{(\text{trace}(XX^{\top}XX^{\top})) \times (\text{trace}(YY^{\top}YY^{\top}))}}
\label{RVcoef}
\end{equation*}
\noindent and takes values in $[0,1]$, with values closer to $1$ indicating higher similarity. The results are shown in  Figure~\ref{fig:SNR2_Boxplots_loadings}. For Scenarios~A and B, the RV values of the corresponding shared and group-specific factor loadings range from $0.75$ to $1.0$, indicating good similarity between the true and estimated matrices.  Notably, in Scenario~C, even though the method's performance in recovering the correct factor configurations deteriorates considerably, as shown in Table \ref{tab:SNR2_esti_num_factors}, once the true configuration is achieved the group-specific factor loading structures are nearly perfectly estimated.

In the Supplementary Materials we report results for additional simulated scenarios with $S=3$ groups and $L=3$ and $(K_1,K_2,K_3)=(2,2,2)$, considering both balanced and unbalanced sample sizes across groups, under $\text{SNR}=2$ and $\text{SNR}=5$. Findings are consistent with those reported above for $S=2$ groups and Scenario A $(3,2,2)$. We note that as the number of groups increases, the computational burden grows since each additional group introduces its own factor loadings and scores, and the corresponding Gibbs updates require repeated inversion and multiplication of high-dimensional matrices involving Kronecker-product terms.  With our implementation of the model, one MCMC chain took about 1 hour to run for the $S=2$ scenarios and about 1.25 to 1.9 hours for the cases with $S=3$.

\subsection{Comparative Study}
Next we perform a comparative study to assess the performances of our BMGFFM method against competing methods. We consider two Bayesian approaches: the semiparametric functional factor model (SFFM) of \cite{KowalandCanale2023} and the Bayesian functional PCA (BFPCA) approach of \cite{Bolandetal2023}.
The SFFM model includes a parametric part, specified through a chosen template (e.g., linear, linear change, cosinor, biphasic) in the original code provided by the authors. In our simulation study, however, there is no comparable parametric structure that should naturally be captured by such template. To help with this, we consider a modified implementation, replacing the linear parametric specification with a simpler intercept-only specification, together with functional centering and a larger initial dimension for the nonparametric component ($K=20$), keeping the default values of the parameter-expanded shrinkage prior, as suggested by the authors. As for the Bayesian functional PCA (BFPCA) approach of \cite{Bolandetal2023}, we use a fixed number of factors specified as $\max(6,[R/4])$, where $R$ denotes the number of basis functions, and default values for their modified MGP prior, as suggested by the authors. Both methods are applied separately to each group, to obtain group-specific estimates. Additionally, we consider two classical approaches: the standard fPCA implemented using the \texttt{fda} package in R \citep{Ramsayetal2022}, with the number of retained eigencomponents chosen to explain at least $90\%$ of the total variation, and the filtrated common principal component analysis method (filt-fPCA) of \cite{Jiao2024filtrated}, which employs a tree-structured approach for multi-group functional data to extract both global and idiosyncratic components via a multi-layer representation. In filt-fPCA the number of layers $D$ corresponds to the depth of the hierarchical decomposition. We set $D=5$ in all scenarios, to achieve a value close to $90\%$ of explained covariance. 

\begin{figure}
    \centering
    \includegraphics[width=0.9\linewidth]{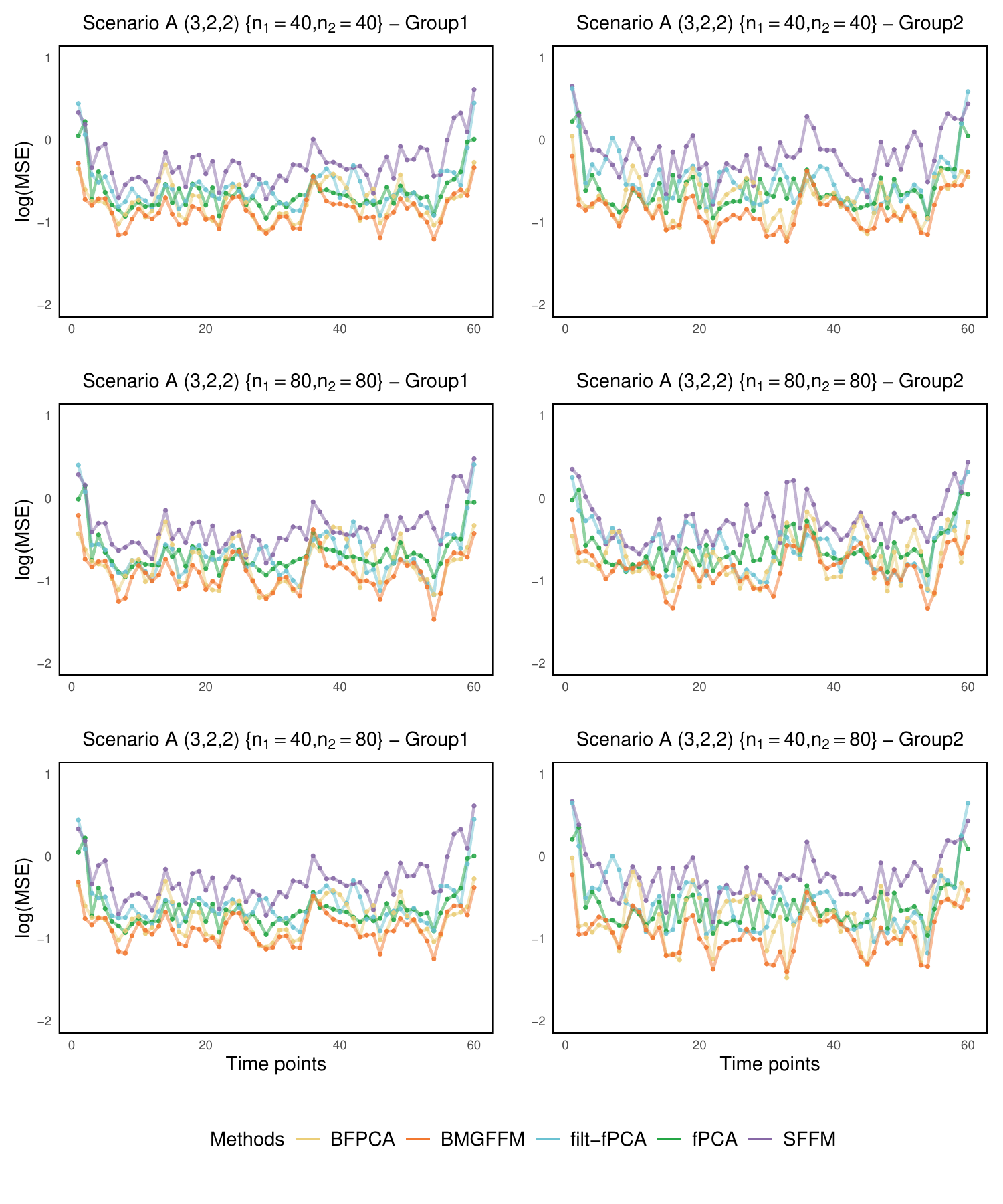}
    \caption{\textbf{Simulation Study}. Group-level point-wise $\log(\mathrm{MSE})$, averaged across 100 replicated datasets, for scenario~A~$(3,2,2)$ with different sample sizes.
We compare our method, BMGFFM, with the semiparametric functional factor model (SFFM) of \cite{KowalandCanale2023}, the Bayesian functional PCA (BFPCA) approach of \cite{Bolandetal2023}, the standard fPCA implemented using the \texttt{fda} package in R \citep{Ramsayetal2022}, and the filtrated common principal component analysis method (filt-fPCA) of \cite{Jiao2024filtrated}. }
\label{fig:SNR2_MSE_competitor_plotssceA1_2_3}
\end{figure}

In order to compare estimation performances, we calculate point-wise mean squared errors (MSEs) between the true individual trajectories $\boldsymbol{f}_{is}$ and their estimated counterparts $\widehat{\boldsymbol{f}}_{is}$ at the group level, averaged over all subjects and all 100 simulated datasets, as 
\begin{equation}
\overline{\text{MSE}}_{s}(t) = \frac{1}{100 \times n_{s}}\sum_{j=1}^{100} \sum_{i=1}^{n_{s}}\left(f_{ijs}(t) - \widehat{f}_{ijs}(t)\right)^{2}, 
\qquad t\in\{t_1,\ldots,t_T\},\end{equation}
where $n_{s}$ is the number of subjects in group $s$. We also calculate the total MSE for each group as $\overline{\text{MSE}}_{s} = \frac{1}{T}\sum_{t=1}^{T}\overline{\text{MSE}}_{s}(t)$. Table \ref{RV_MSE} reports the total MSE values for all scenarios and methods considered. BMGFFM consistently achieves the lowest MSE in Scenario~A~$(3,2,2)$, where both shared and group-specific latent structures are present in both groups. In scenarios where group-specific structure is present only in some groups, such as Scenario~A~$(3,2,0)$, BMGFFM attains better estimation performance in group 1, which exhibits group-specific variation, whereas BFPCA performs slightly better in the simpler group (group 2) with only shared variation. Nevertheless, the differences in MSE between the two methods remain small, indicating that BMGFFM can achieve optimal estimation when groups contain both shared and group-specific functional patterns. In Scenario~B, where only the shared structure is present, BFPCA achieves the lowest MSE. The performance of BMGFFM is comparable, indicating that both models capture the shared variations with similar accuracy. In Scenario~C, which contains only group-specific structures, BFPCA yields the lowest MSE, whereas BMGFFM and fPCA still achieve reasonable reconstruction accuracy and performs substantially better than filt-fPCA and SFFM. 
Point-wise MSE curves for Scenario A $(3, 2, 2)$ with different sample sizes, plotted on the log scale for ease of visualization, are shown in Figure~\ref{fig:SNR2_MSE_competitor_plotssceA1_2_3}. BMGFFM consistently yields lower and more stable errors across time in both groups, particularly at time points where alternative methods tend to have higher errors. While standard fPCA and BFPCA perform reasonably well, SFFM and filt-fPCA exhibit larger errors, especially at the beginning and end of the observed interval, likely due to limited information for estimating the curves. Point-wise log(MSE) curves for Scenario~A~$(3,2,0)$ and those for Scenarios~B~$(3,0,0)$ and C~$(0,2,2)$ are provided in the Supplementary Materials. 
Overall, performance varies across scenarios and depends on the underlying latent structure. BMGFFM is particularly well suited to settings where both shared and group-specific functional variation are present, which is consistent with its modeling goals. 
By contrast, in simpler settings dominated by a single type of latent variation, BFPCA becomes more competitive and, in Scenarios~B~$(3,0,0)$ and C~$(0,2,2)$, often attains slightly lower errors than BMGFFM.

\begin{table}[!htbp]
\centering
\caption{\textbf{Simulation Study}. Total MSEs, averaged across 100 replicated datasets, for all simulated scenarios described in Section \ref{subsec:gatagen}.
We compare our method, BMGFFM, with the semiparametric functional factor model (SFFM) of \cite{KowalandCanale2023}, the Bayesian functional PCA (BFPCA) approach of \cite{Bolandetal2023}, the standard fPCA implemented using the \texttt{fda} package in R \citep{Ramsayetal2022}, and the filtrated common principal component analysis method (filt-fPCA) of \cite{Jiao2024filtrated}. Smallest MSE values are in bold.} 
\scriptsize
\begin{tabular}{c c c c c c c c c c c c c}
\Xhline{1pt}
\multirow{2}{*}{Sce.} & \multirow{2}{*}{($L,K_1,K_2$)} & \multirow{2}{*}{$\{n_1,n_2\}$} 
& \multicolumn{2}{c}{BFPCA} 
& \multicolumn{2}{c}{BMGFFM}
& \multicolumn{2}{c}{filt-fPCA} 
& \multicolumn{2}{c}{fPCA} 
& \multicolumn{2}{c}{SFFM} \\
\cmidrule(lr){4-5} \cmidrule(lr){6-7} \cmidrule(lr){8-9} \cmidrule(lr){10-11} \cmidrule(lr){12-13}
 & & & $\overline{\text{MSE}}_{1}$ & $\overline{\text{MSE}}_{2}$ & $\overline{\text{MSE}}_{1}$ & $\overline{\text{MSE}}_{2}$ & $\overline{\text{MSE}}_{1}$ & $\overline{\text{MSE}}_{2}$ & $\overline{\text{MSE}}_{1}$ & $\overline{\text{MSE}}_{2}$ & $\overline{\text{MSE}}_{1}$ & $\overline{\text{MSE}}_{2}$ \\
\addlinespace[1mm]
\hline
\addlinespace[1mm]
A & (3,2,2) & \{40,40\} & 0.20 & 0.22 & \bf{0.15} & \bf{0.16} & 0.36 & 0.49 & 0.28 & 0.33 & 0.65 & 0.85 \\
A & (3,2,2) & \{80,80\} & 0.19 & 0.22 & \bf{0.14} & \bf{0.17}  & 0.31 & 0.34 & 0.26 & 0.30 & 0.55 & 0.66 \\
A & (3,2,2) & \{40,80\} & 0.20 & 0.23 & \bf{0.15} & \bf{0.14} & 0.36 & 0.46 & 0.28 & 0.33 & 0.65 & 0.74 \\
A & (3,2,0) & \{40,40\} & 0.20 & \bf{0.06} & \bf{0.15}  & 0.07 & 0.35 & 0.30 & 0.28 & 0.18 & 0.65 & 0.48 \\
A & (3,2,0) & \{80,80\} & 0.19 &  \bf{0.04} & \bf{0.15}  & 0.07 & 0.31 & 0.30 & 0.26 & 0.19 & 0.55 & 0.45 \\
A & (3,2,0) & \{40,80\} & 0.20 &  \bf{0.04} & \bf{0.17}  & 0.08 & 0.34 & 0.29 & 0.28 & 0.19 & 0.65 & 0.43 \\
\addlinespace[1mm]
\hline
\addlinespace[1mm]
B & (3,0,0) & \{40,40\} & {\bf 0.06} & {\bf 0.06} &  0.08 & 0.08  & 0.23 & 0.23 & 0.18 & 0.19 & 0.49 & 0.48 \\
B & (3,0,0) & \{80,80\} & {\bf 0.04} & {\bf 0.04} &  0.07 & 0.07  & 0.19 & 0.20 & 0.17 & 0.19 & 0.41 & 0.41 \\
B & (3,0,0) & \{40,80\} & {\bf 0.06} & {\bf 0.04} & {\bf 0.06} & 0.06  & 0.22 & 0.24 & 0.18 & 0.19 & 0.49 & 0.44 \\
\addlinespace[1mm]
\hline
\addlinespace[1mm]
C & (0,2,2) & \{40,40\} & {\bf 0.03} & {\bf 0.03} & 0.05 & 0.05  & 0.21 & 0.19 & 0.13 & 0.09 & 0.34 & 0.17 \\
C & (0,2,2) & \{80,80\} & {\bf 0.02} & {\bf 0.02} & 0.05 & 0.05  & 0.20 & 0.20 & 0.14 & 0.10 & 0.31 & 0.24 \\
C & (0,2,2) & \{40,80\} & {\bf 0.03} & {\bf 0.02} &  0.05 & 0.05  & 0.21 & 0.18 & 0.13 & 0.09 & 0.34 & 0.15 \\

\Xhline{1pt}
\end{tabular}
\label{RV_MSE}
\end{table}

\begin{figure}
    \centering
    \includegraphics[width=0.97\linewidth]{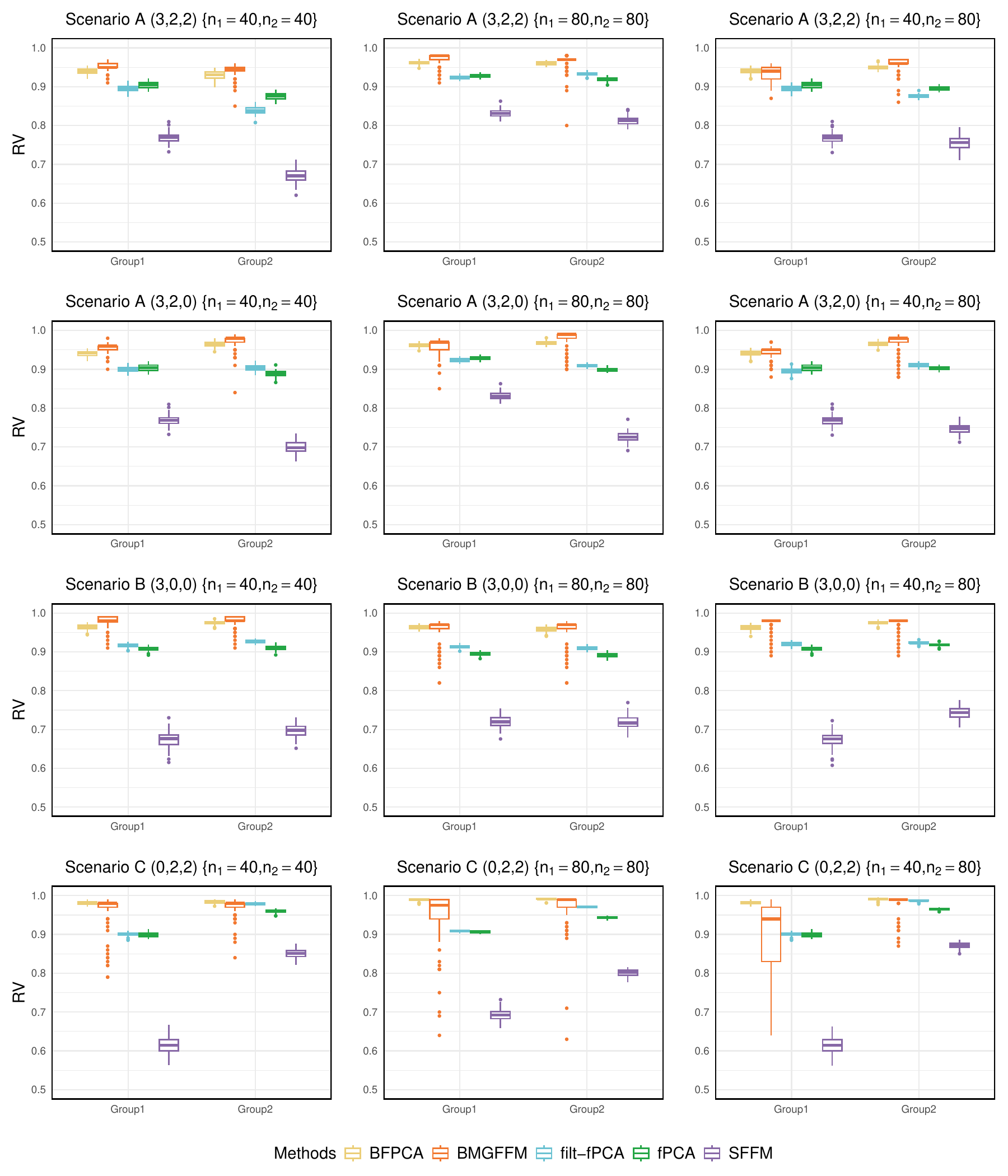}
    \caption{\textbf{Simulation Study}. Boxplots of RV coefficients between true and estimated explained covariance matrices, across 100 replicates, under each simulated scenario. We compare our method, BMGFFM, with the semiparametric functional factor model (SFFM) of \cite{KowalandCanale2023}, the Bayesian functional PCA (BFPCA) approach of \cite{Bolandetal2023}, the standard fPCA implemented using the \texttt{fda} package in R \citep{Ramsayetal2022}, and the filtrated common principal component analysis method (filt-fPCA) of \cite{Jiao2024filtrated}.} 
    \label{fig:SNR2_comp_boxplots_A1-C3}
\end{figure}

Finally, we show the RV coefficients between the estimated and true explained covariance matrices, across 100 replicated datasets, for each scenario. Boxplots of the estimates are reported in Figure~\ref{fig:SNR2_comp_boxplots_A1-C3}. 
Again, across scenarios the relative performance depends on the underlying latent structure. In Scenarios~A~$(3,2,2)$, A~$(3,2,0)$, and B~$(3,0,0)$, BMGFFM and BFPCA generally achieve highest RV coefficients for both groups, thought some outliers are noticeable for BMGFFM. In these scenarios, filt-fPCA and fPCA typically yield slightly lower RV coefficients, while SFFM tends to produce even lower values. In Scenario~C~$(0,2,2)$, however, BFPCA tends to achieve the highest RV coefficients. Since there is no shared latent structure in this setting, applying BFPCA separately within each group is more closely aligned with the data-generating mechanism.

\subsection{Sensitivity Analysis}
\label{sensitivity analysis}
We focus on Scenario~A~$(3,2,2)$ with $n_1=40$, $n_2=80$ and investigate the sensitivity of the proposed model to the choice of hyperparameters $a_{1}, a_{2}$ and $v_{0}$ in the prior on both the shared and group-specific factor loading matrices. We consider values $(a_{1},a_{2}) \in \{(5,25), (5,50), (10,30)\}$ and $v_{0} \in \{0.01, 0.001, 0.005, 0.00025\}$, following the settings in \cite{Scheipletal2012}. For consistency, we adopt the same set of hyperparameters for both shared and group-specific terms, ensuring a comparable level of shrinkage. Table~\ref{tab: SNR2_sensitivity analysis} reports results as RV values between true and estimated explained matrices and total MSEs between true and estimated curves, averaged over $50$ replicates for each hyperparameter combination.
\begin{table}[ht]
\centering
\caption{\textbf{Simulation Study}. Sensitivity analysis for hyperparameters $(a_1, a_2, v_0)$ based on $50$ replicated datasets of Scenario~A~$(3,2,2)$ with $n_{1}=40$, $n_{2}=80$. Results are mean RV values, with standard deviation (in parentheses), between true and estimated explained matrices $\Sigma_{f_s}$, and total MSE between true and estimated curves, averaged across subjects, time points and replicates.}
\begin{tabular}{c c c c c}
\Xhline{1pt}
$(a_1, a_2, v_0)$ & $\Sigma_{f_1}$ & $\Sigma_{f_2}$ & $\overline{\text{MSE}}_{1}$ & $\overline{\text{MSE}}_{2}$ \\ 
\addlinespace[1mm]
\hline
\addlinespace[1mm]
(5, 25, 0.00025) & 0.94 (0.02) & 0.96 (0.01) & 0.15 & 0.14  \\ 
(5, 25, 0.001) & 0.93 (0.03) & 0.95 (0.03) & 0.16 & 0.16  \\ 
(5, 25, 0.005) & 0.92 (0.03) & 0.95 (0.03) & 0.20 & 0.21  \\ 
(5, 25, 0.01) & 0.82 (0.09) & 0.82 (0.12) & 0.75 & 0.88  \\ 
\addlinespace[1mm]
\hline
\addlinespace[1mm]
(5, 50, 0.00025) & 0.93 (0.02) & 0.96 (0.02) & 0.15 & 0.16  \\ 
(5, 50, 0.001)  & 0.94 (0.01) & 0.95 (0.03) & 0.17 & 0.19  \\
(5, 50, 0.005)  & 0.79 (0.11) & 0.77 (0.11) & 1.07 & 1.23  \\ 
(5, 50, 0.01)  & 0.70 (0.07) & 0.70 (0.09) & 1.60 & 1.80  \\ 
\addlinespace[1mm]
\hline
\addlinespace[1mm]
(10, 30, 0.00025) & 0.94 (0.02) & 0.96 (0.01) & 0.14 & 0.13  \\ 
(10, 30, 0.001)  & 0.94 (0.02) & 0.96 (0.02) & 0.15 & 0.16  \\ 
(10, 30, 0.005)  & 0.94 (0.02) & 0.96 (0.02) & 0.16 & 0.15  \\ 
(10, 30, 0.01)  & 0.93 (0.02) & 0.95 (0.02) & 0.21 & 0.20  \\ 
\Xhline{1pt}
\end{tabular}
\label{tab: SNR2_sensitivity analysis}
\end{table}

The results indicate that the number of active factors is more sensitive to the choice of $v_0$ than to $(a_1,a_2)$. The parameter $v_0$ controls the threshold for selecting active factors: in general, smaller values of $v_0$ allow the model to include more factors that explain small variations, while larger values of $v_0$ lead to a more conservative selection. As shown in Table~\ref{tab: SNR2_sensitivity analysis}, when $v_{0}$ is small ($0.00025$ or $0.001$), the model recovers the underlying covariance structures well, with RV values above $0.93$ and relatively low MSEs. In contrast, larger values of $v_{0}$ (0.005 or 0.01) result in a clear drop in RV values and higher estimation errors between true and estimated curves. The best performance is obtained for $(5,25,0.00025)$ which corresponds to the default prior setting of \cite{Scheipletal2012}, with settings $(10,30,0.00025)$ and $(10,30,0.001)$ yielding comparable results.
Among these settings, $(10,30,0.00025)$ and $(5,25,0.00025)$ impose less shrinkage a priori and tend to include more active factors, which may result in the inclusion of weaker factors when the data are less complex. To achieve more conservative results across different datasets, we therefore adopt $(10,30,0.001)$ as the default setting in all applications of this paper.

%---------------------------------------------------
\section{Application to Event-Related Potentials}
\label{sec:Application}
For a real data analysis, we consider EEG data on alcoholic and healthy controls, originally collected as part of the COGA (Collaborative Studies on the Genetics of Alcoholism) project, and available from the UCI Machine Learning Repository \citep{BacheandLichman2013}.

\subsection{EEG Data}
Details and background information on the experimental study can be found in \cite{Zhangetal1995} and \cite{Ingber1997statistical}. 
The full dataset includes EEG measurements from $77$ alcoholic and $45$ control subjects, however data from subjects $co2a0000425$ and $co2c0000391$ were excluded due to their small amounts of data, as noted by \cite{HelwigandPign2016}. 
Each subject was exposed to three different experimental conditions involving one or two visual stimuli,
with images selected from the picture set of \cite{SnodgrassandVanderwart1980}.
EEG recordings were obtained from 64 electrodes placed at standard locations on the subjects' scalps, sampled at 256 Hz ($3.9$-ms epoch) for 1 second. 
As noted in \cite{Qazietal2021}, $19$ electrodes are particularly relevant for alcoholism detection and medical diagnosis: 7 frontal (FP1, FP2, F8, AF7, AF8, F4, F3), 5 central (FC6, FC5, FC2, C3, C4), 2 parietal (CP2, PZ), 2 occipital (PO2, PO1), and 3 temporal (T8, CP6, P7).  
In our analysis, for each subject we focused on the experimental condition with a single stimulus and averaged the EEG measurements across the $19$ electrodes and the available trials, to obtain smooth event-related potentials (ERPs) signals. 
Results we report below were obtained by considering the time window from $47$ms to $349$ms, comprising $78$ common grid points, that captures key ERP components relevant to cognitive and perceptual processing. Specifically, this time window includes the N100 and P200 components, which are commonly studied in EEG research related to attention allocation and perception \citep{Proverbioetal2022, Qazietal2021}. 
The subject-level ERP curves are shown in the top panel of Figure \ref{fig:2groups}, for the two subject groups. 

\subsection{Parameter Settings}
The hyperparameters were set to the same values used in our simulation study, that is $a_{1}=10, a_{2}=30,$ and $v_{0}=0.001$ for the shared factors and $a^{s}_{1}=10, a^{s}_{2}=30,$ and $v^{s}_{0}=0.001$ for the group-specific factors. We used $R=40$ B-spline basis functions, and ran the MCMC algorithm for $20,000$ iterations, discarding the first $10,000$ as burn-in.  As with the simulation study, we assessed convergence by calculating the Geweke's convergence diagnostic \citep{Geweke1992evaluating} for the model parameters and by visually inspecting traceplots. Both $\sigma^{2}_{\epsilon_{s}}$ and $\sigma^{2}_{\beta_{s}}$ showed Geweke statistics well within $\pm 1.96$ and traceplots showed no drift with good mixing (see Supplementary Materials).
Furthermore, inspection of representative elements of $\beta_{s}$ also revealed acceptable diagnostics and stable behavior (see Supplementary Materials). Overall, these diagnostics indicate that the MCMC sampler convergence adequately across key components of the model, providing reliable posterior inference.

\subsection{Results}
\begin{figure}[htbp]
    \centering
    \includegraphics[width=.8\linewidth]{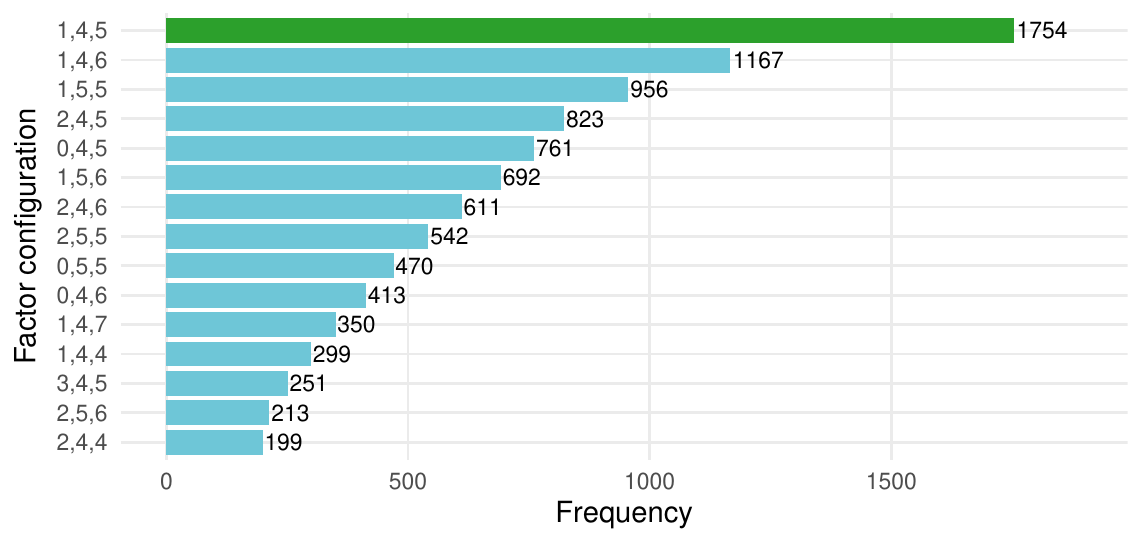}
    \caption{\textbf{Application to Event-Related Potentials.} Posterior frequencies of the $15$ most frequently sampled factor configurations across $10,000$ retained MCMC draws after burn-in.}
    \label{fig:EEG_histogram_factor_configuration}
\end{figure}

\begin{figure}[htbp]
    \centering
\includegraphics[width=1\textwidth]{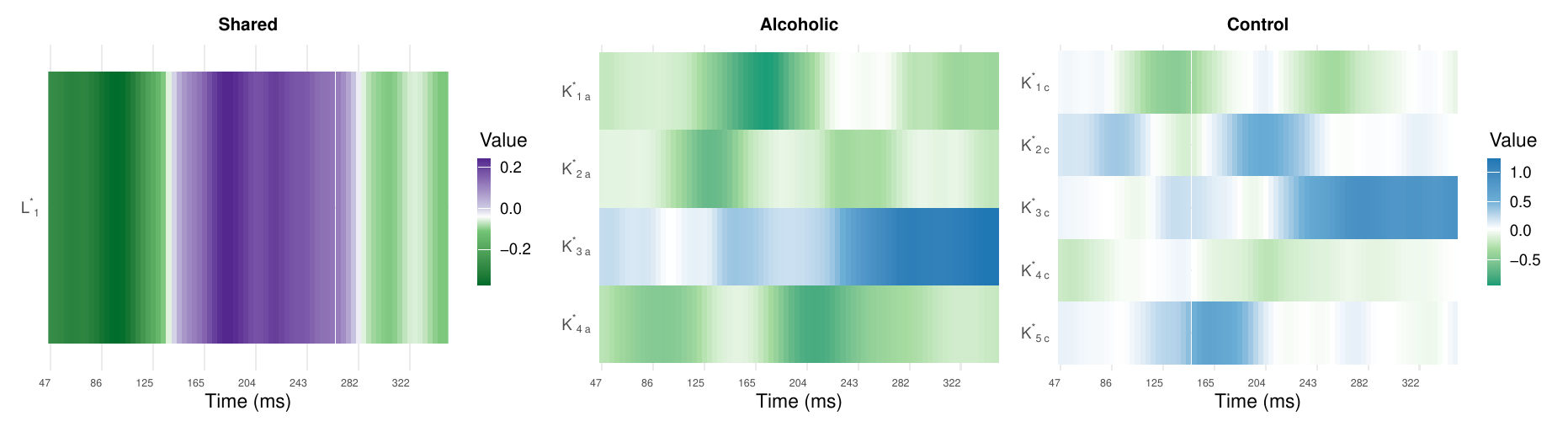}
\caption{\textbf{Application to Event-Related Potentials}. Heatmaps of the estimated covariance-derived shared and group-specific factor loadings, over a time window from $47$ms to $349$ms. The shared component $\widetilde{\Lambda}^{\top}$, of dimension $1 \times 78$, is shown on the left-hand side, the alcoholic-specific component $ \widetilde{\Phi}^{\top}_{\text{alcoholic}}$, of dimension $4 \times 78$, is shown in middle panel, and the control-specific component $\widetilde{\Phi}^{\top}_{\text{control}}$, of dimension $5 \times 78$, in the right panel.} 
\label{fig:real_data_loadings}
\end{figure}

Figure~\ref{fig:EEG_histogram_factor_configuration} shows the posterior distribution of the factor configurations across $10,000$ retained MCMC draws after burn-in, restricted to the 15 most frequently sampled factor configurations.  The most frequently sampled combination is 
$(L^{*}=1, K^{*}_{\text{alcoholic}}=4, K^{*}_{\text{control}}=5)$, 
which we adopt as the optimal choice. Accordingly, the model identified one shared factor, four alcoholic-specific factors, and five control-specific factors. We expected the shared structure to capture neural patterns common across the two groups, while the group-specific structure would reveal differential activity unique to each group. Accordingly, as shown in Figure~\ref{fig:real_data_loadings}, the shared factor loadings exhibit smooth, low-amplitude fluctuations across the entire time windows. The shared factor loadings specifically resemble well-known ERP components: the negative deflection around 100ms is consistent with an N100-like response, while the positive deflection between 
$160-250$ms aligns with P200 activity, both reflecting fundamental sensory and perceptual processes that likely represent background activity rather than strongly localized variations. In contrast, the group-specific factor loadings reveal distinct temporal heterogeneity. For the alcoholic group, the third factor shows more pronounced positive fluctuations between 
$240-347$ms, which may correspond to altered or delayed neural responses compared with the control group. Other factors display broader fluctuations over time, consistent with less temporally focused engagement. For the control group, the third factor also shows positive fluctuations between $240-347$ms, although slightly weaker than in the alcoholic group. The remaining factor loadings tend to be more temporally localized, which is consistent with more synchronized neural engagement during stimulus evaluation.

\begin{figure}[ht]
    \centering
    \includegraphics[width=0.8\linewidth]{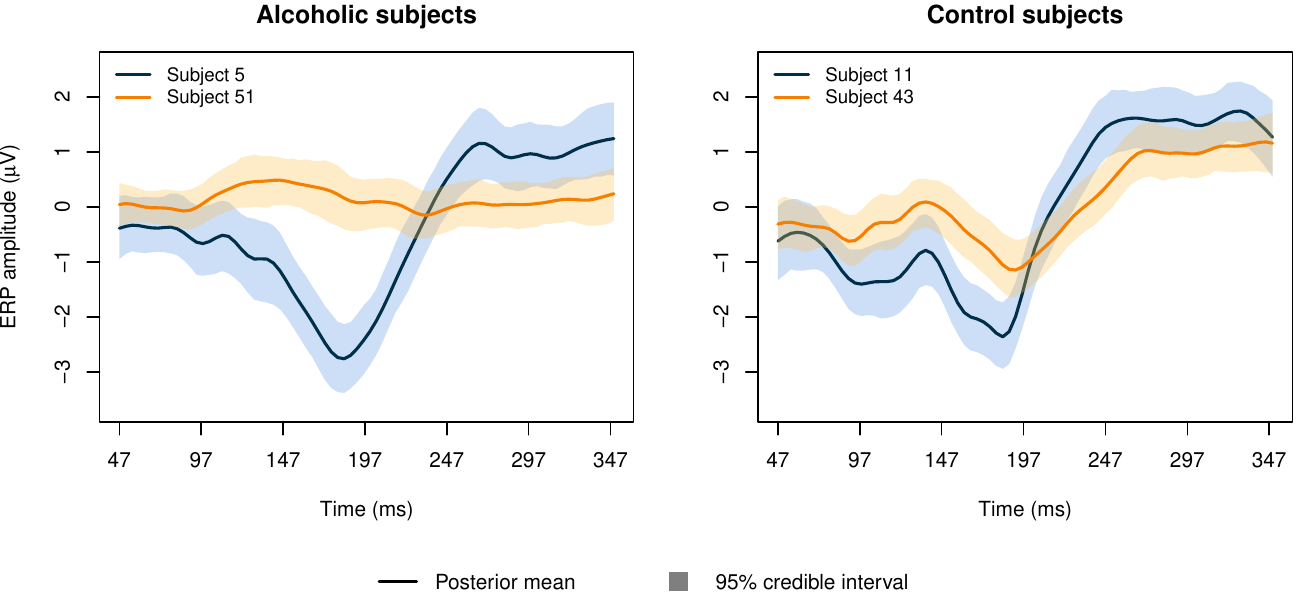}
    \caption{\textbf{Application to Event-Related Potentials.} Posterior mean estimates and pointwise $95\%$ credible intervals for two representative subjects from the alcoholic group and two representative subjects from the control group, based on posterior draws corresponding to the modal factor configuration $(L^{*},K_{a}^{*},K_{c}^{*})=(1,4,5)$.}
\label{fig:realdata_representative_subjects_two_panels}
\end{figure}

\begin{figure}[!ht]
    \centering
\includegraphics[width=0.8\textwidth]{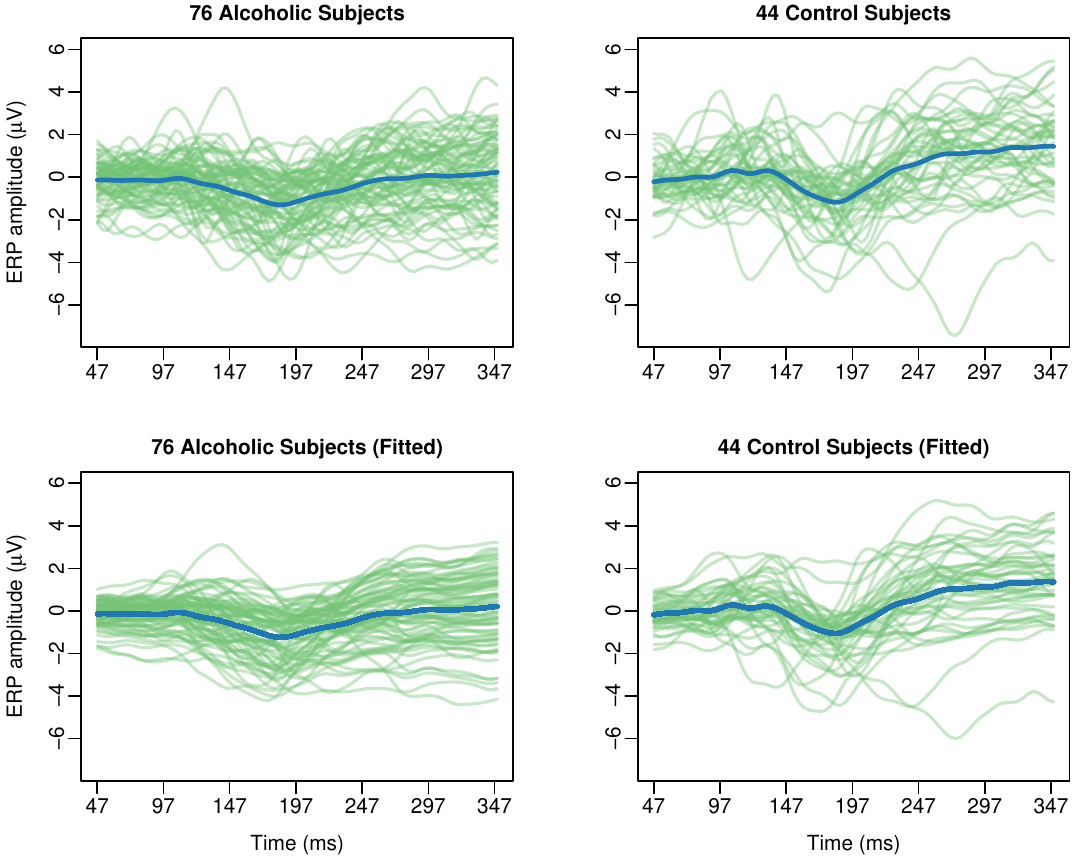}
\caption{\textbf{Application to Event-Related Potentials}. Top line: Event-Related Potentials (ERPs) for 76 alcoholic subjects (left) and 44 control subjects (right) over the time window from $47$ ms to $349$ ms. The average ERP curves are indicated as a green curve. Bottom line: fitted curves as posterior means from BMGFFM, with group-specific mean function shown in blue.} 
\label{fig:2groups}
\end{figure}

Figure~\ref{fig:realdata_representative_subjects_two_panels} reports estimated ERP curves and pointwise $95\%$ credible intervals for two representative subjects from the alcoholic group and two representative subjects from the control group, based on posterior draws corresponding to the modal factor configuration $(L^{*},K_{a}^{*},K_{c}^{*})=(1,4,5)$, while Figure \ref{fig:2groups} shows the estimated ERP curves for all subjects in the two groups, together with the group-level mean functions. The estimated curves for the control subjects exhibit clearer and more differentiated peaks between approximately 140 ms and 250 ms after stimulus onset, while those of the alcoholic subjects show delayed and blurred responses in this window, with less pronounced differences between stimuli. Taken together, these results demonstrate that our model provides an effective representation of underlying neural dynamics, capturing both common and distinct features of cognitive processing.

%---------------------------------------------------
\FloatBarrier
\section{Conclusion}
\label{sec:Conclusion}
In this paper, we have proposed a Bayesian factor model for multi-group functional data that uses an explicit decomposition of the data into group-specific mean functions and latent components that capture both common and distinct latent structures across the groups. Our approach uses a common set of cubic B-spline basis functions to achieve a low-rank representation of the latent factor functions.  Sparsity is promoted by placing a parameter-expanded cumulative shrinkage process prior on the factor loadings, which induces increasing shrinkage while allowing inference on the number of active shared and group-specific factors. We have carried out posterior inference via Gibbs sampler, coupled with a post-processing strategy to address identifiability issues.
Using simulated data, we have demonstrated that the proposed model accurately recovers the underlying latent factor structures across diverse scenarios, improving upon competing methods. In an application to EEG data, our model has identified both shared brain responses associated with known event-related potentials and group-specific neural dynamics distinguishing alcoholic and control subjects.

In our current implementation, the group-specific latent components are modeled separately for each group, and no additional dependence structure is imposed across groups beyond the shared component. The assumption of no additional dependence across groups beyond the shared structure is consistent with our real EEG application, where the two groups correspond to alcoholic and control subjects and the observations arise from distinct individuals in each group. If dependence were present among the group-specific components, estimation would likely become more challenging, because variation common to only a subset of groups could be harder to distinguish from the globally shared structure. Future work could extend the model to accommodate factors shared among subsets of groups, as in \cite{Grabskietal2023} and \cite{BortolatoandCanale2026}.

\bibliographystyle{apalike}
\bibliography{references}

\end{document}